\documentclass[twocolumn]{aastex61}

\usepackage{subfigure}
\usepackage{amsmath}
\usepackage{url}

\received{}
\revised{}
\accepted{}
\submitjournal{ApJ}

\shorttitle{From UFOs to molecular outflows}
\shortauthors{Mizumoto et al.}

\begin{document}

\title{Kinetic energy transfer from X-ray ultrafast outflows to mm/sub-mm cold molecular outflows in Seyfert galaxies}

\correspondingauthor{MM}
\email{misaki.mizumoto@durham.ac.uk, mizumoto.misaki@gmail.com}

\author{Misaki Mizumoto}
\affil{Centre for Extragalactic Astronomy, Department of Physics, University of Durham, South Road, Durham DH1 3LE, UK}

\author{Takuma Izumi}
\affiliation{National Astronomical Observatory of Japan, Osawa, Mitaka, Tokyo 181-8588, Japan}
\affiliation{NAOJ fellow}

\author{Kotaro Kohno}
\affiliation{Institute of Astronomy, Graduate School of Science, The University of Tokyo, Osawa, Mitaka, Tokyo 181-0015, Japan}

\begin{abstract}
UltraFast Outflows (UFOs), seen as X-ray blueshifted absorption lines in active galactic nuclei (AGNs), are considered to be a key mechanism for AGN feedback.
In this scenario, UFO kinetic energy is transferred into the cold and extended molecular outflow observed at the mm/sub-mm wavelength, which blows away the gas and suppresses star formation and accretion onto the central black hole (BH).
However, the energy transfer between the inner UFO and the outer molecular outflow has not yet fully studied mainly due to the limited sample.
In this paper, we performed comparison of their kinetic energy using the mm/sub-mm published data and the X-ray archival data.
Among fourteen Seyfert galaxies whose molecular outflows are detected in the IRAM/PdBI data,
eight targets are bright enough to perform spectral fitting in X-ray,
and we have detected UFO absorption lines in six targets with 90\% significance level, using XMM-Newton and Suzaku satellites.
The time-averaged UFO kinetic energy was derived from the spectral fitting.
As a result, we have found that the energy-transfer rate (kinetic energy ratio of the molecular outflow to the UFO) ranges from $\sim7\times10^{-3}$ to $\sim$1, and has a negative correlation with the BH mass, which shows that 
the AGN feedback is more efficient in the lower mass BHs.
This tendency is consistent with the theoretical prediction that the cooling time scale of the outflowing gas becomes longer than the flow time scale when the BH mass is smaller.
\end{abstract}


\section{Introduction}

A strong correlation has been observationally established between the mass of a super massive black hole (SMBH) and the physical parameters of its host galaxy, e.g. velocity dispersion of the bulge (\citealt{kor13} for a review).
Currently, the most promising model to explain it is a ``co-evolution'' scenario, that is,
evolutions of active galactic nuclei (AGNs) and their host galaxies control each other.
However, the exact mechanism is still under debate. 
The typical radius of the gravitational field of a BH is much smaller than the size of its host galaxy.
The host galaxy is thus not disturbed by the central SMBH with gravitation but with some other form of interaction, which is energetic enough to yield a significant correlation in the physical parameters between them (\citealt{kin15} for a review).
The ultrafast outflow (UFO) is one of the plausible  interactions that may trigger the co-evolution.
The UFO is a fast ($\sim0.1-0.3\,c$) and ionized wind emanating from  a close vicinity  ($\sim10^{-2}$~pc) of a SMBH.
So far, X-ray spectroscopic observations have shown that about half of AGNs have UFOs, which make blueshifted iron absorption lines in their X-ray energy spectra (e.g., \citealt{tom10}).
UFOs are mainly seen in the super-Eddington sources (e.g. \citealt{kin03b}), 
but detected even in low $L/L_\mathrm{Edd}$ AGNs, where $L$ is  the bolometric luminosity of an AGN and $L_\mathrm{Edd}$ is the Eddington luminosity.
Their fast velocity and wide solid angle ($\Omega/2\pi\simeq0.4$; \citealt{gof15}) enable it to transport a significant amount of kinetic energy from an AGN to its host galaxy.
In this way, UFOs are considered to affect the co-evolution.
This type of feedback is called  the ``quasar mode'', in contrast to  the ``radio mode'', where highly collimated jets take away the kinetic energy.
However, observational evidence that UFOs contribute to the co-evolution  is still  very limited, mainly because
X-ray observations can trace only the close vicinity of  central SMBHs.
Therefore, we  need to use other wavelengths to constrain the UFO contribution to the galaxy-scale gas.

The kinetic energies of UFOs are theoretically considered to be transferred into the molecular outflows \citep{zub12,pou13}.
The molecular outflow is a cold(est) gas outflow observed in many AGNs at the mm/submm and far-infrared  wavelengths with a size of $\sim$400~pc and  velocity of $\sim500$~km~s$^{-1}$ (e.g., \citealt{cic14}).
It is considered to be an accumulation of the ambient gas swept by the shock fronts that UFOs have created \citep{kin15}.
The molecular gas is responsible for the gas mass in the central region of the galaxy ($\lesssim1$~kpc; see Fig.\,2 in \citealt{hon95}), so that the molecular outflows play a dominant role to carry the kinetic energies.
Observational clues have also been shown that the molecular outflow reaches the circumnuclear disk, quenches star formation, and contributes to the AGN feedback (e.g., \citealt{fer10,gar14}).

Here, the following question is raised:
``Is the kinetic energy of UFOs efficiently transferred to the molecular outflows?''
Many studies (implicitly or explicitly) assume that UFO energy is mechanically carried to the molecular outflow and drives the galactic-scale feedback (e.g., \citealt{cap09,gof13,gof15,ham18}).
However, the UFO energy is lost via radiative cooling before the shock front \citep{pou13};
If radiative cooling is efficient, the UFO loses its energy before reaching the shock front and has little contribution to the co-evolution.
In this situation, the UFO creates a ``momentum-conserving shock''.
If not, UFO's energy is directly transferred to the host galaxy and the UFO produces a ``energy-conserving shock''.
To evaluate whether the UFO kinetic energy is carried to the molecular outflow effectively, we introduce the ``energy-transfer rate'' $C=\dot{K}_{\rm molecular}/\dot{K}_{\rm UFO}$, where $K_i$ is the kinetic energy rate of molecular outflows and UFOs.
On one hand, if $C\sim1$ (i.e., energy-conserving shock), the UFO energy can flow into the large-scale molecular outflow and affect the AGN feedback.
On the other hand, if $C\ll1$ (i.e., momentum-conserving shock), most of the UFO energy is lost and cannot contribute to quench star formation.
Therefore, measuring the energy-transfer rate with large samples is essential to investigate validity of the quasar-mode feedback scenario.

Attempts to compare the two types of outflows and investigate energy transfer have been performed,
but the samples are very limited.
\citet{tom15} reported a powerful X-ray UFO in IRAS F11119+3257 and investigated kinetic energy transformation to the large-scale molecular outflows (also see \citealt{vei17}).
\citet{fer17} compared outflow parameters of three targets, IRAS F11119+3257, Mrk 231, APM 08279+5255, and argued that types of shock between two outflows are not unique among the three targets.
In this paper, we measure the energy-transfer rate with larger samples.
We use \citet{cic12} as a reference of the molecular outflows, which listed parameters of CO(1--0) molecular outflows of 19 galaxies obtained by {\it Plateau de Bure Interferometer} (PdBI)\footnote{It is now upgraded to the {NOrthern Extended Millimeter Array} (NOEMA).} on Institut de RAdioastronomie Millim{\' e}trique (IRAM), 14 of which are categorized as AGNs.
We search the X-ray archival data of the 14 AGNs, derive their UFO parameters, and compare them with the molecular outflows to calculate the energy-transfer rate.
First, we explain the target selection and X-ray data reduction in \S\ref{sec2}, and
calculate the energy-transfer rates in \S\ref{sec3}.  
Then we discuss under what condition the UFO contribution to the AGN feedback is efficient in \S\ref{sec4}, and finally
give our conclusions in \S\ref{sec5}.

\section{Target selection and data reduction} \label{sec2}

\subsection{Target selection}
We use the X-ray archival data of CCD detectors on {\it XMM-Newton} and {\it Suzaku}, which are most suitable to detect UFO lines with their large effective area.
We found that, among 14 AGNs in \citet{cic12}, 10 targets have been observed in X-rays (table \ref{tab:targets}).
Among them, two targets (IRAS F08572 and IRAS F10565) are too faint to be analyzed.
Therefore, we analyze eight targets to search the UFO lines.

\begin{deluxetable*}{llllcc}[b!]
\tablecaption{X-ray observations of the targets \label{tab:targets}}
\tablecolumns{28}
\tablewidth{0pt}
\tablehead{
\colhead{Object}&\colhead{Name}&\colhead{ID}&\colhead{Date}&\colhead{Duration (s)}&\colhead{Exposure (s)}}
\startdata
IC 5063          & Suzaku  & 704010010     & 2009.4 & ---   & 45160\\
IRAS F08572+3915 & (XMM)     & 0200630101    & 2004.4 & 28918 & 25711\\
                 & (Suzaku)  & 701053010   & 2006.4 & ---   & 77197 \\
IRAS F10565+2448 & (XMM)     & 0150320201    & 2003.6 & 32217 & 22454\\
                 & (Suzaku)  & 702115010     & 2007.11& ---   & 39423 \\
I Zw 1           & XMM1    & 0110890301    & 2002.6 & 21973 & 18176\\
                 & XMM2    & 0300470101    & 2005.7 & 85508 & 57912\\
                 & XMM3    & 0743050301+801& 2015.1 & 275600 & 171243\\
Mrk 231          & XMM1    & 0081340201    & 2001.6 & 22342 & 17205\\
                 & XMM2    & 0770580401+501& 2015.4--5 & 50700 & 39769\\
                 & Suzaku  & 706037010     & 2011.4 & ---   & 197511\\
Mrk 273          & XMM1    & 0101640401    & 2002.5 & 22840 & 17969\\
                 & (XMM2)    & 0601360301--701& 2010.5--6 & 54697 & 151\\
                 & (XMM3)    & 0722610201    & 2013.11& 22800 & 3483\\
                 & Suzaku  & 701050010     & 2006.7 & ---   & 79905\\
Mrk 876          & XMM1    & 0102040601    & 2001.4 & 12825 & 3511\\
                 & XMM2    & 0102041301    & 2001.8 & 7919  & 2593\\
NGC 1068         & XMM1    & 0111200101+201& 2000.7 & 63062 &62985\\
                 & XMM2    & 0740060201--401& 2014.7--8 & 175597 &119095\\
                 & XMM3    & 0740060501    & 2015.2 & 54600 &33851\\
                 & Suzaku  & 701039010     & 2007.2 & ---   & 41623\\
NGC 1266         & XMM     & 0693520101    & 2012.7 & 138580& 81560\\
NGC 6240         & XMM1    & 0101640101    & 2000.9 & 30111 &10119\\
                 & XMM2    & 0101640501-601    & 2002.3 & 18871 & 4763\\
                 & XMM3    & 0147420201& 2003.3 & 31640 & 3050\\
                 & XMM4    & 0147420401-601 & 2003.8 & 54548  &10678 \\
\enddata
\tablecomments{In XMM data, exposure time shows after removing background flares.
We do not perform model fitting of the observations in parentheses, which have too few photon counts to be fitted due to faintness and/or heavy pollution of the background flares.
}
\end{deluxetable*}

\subsection{Data reduction}
In the analysis of the {\it XMM-Newton} data \citep{jan01}, 
we use only the data from the European Photon Imaging Camera (EPIC)-pn \citep{str01}, 
which has the largest effective area around the energy band we focus on.
The data are reduced using the XMM-Newton Software Analysis System ({\tt SAS}, v.15.0.0) and the latest calibration files. 
High background periods, when the count rates in 10--12~keV with {\tt PATTERN==0} are higher than 0.4 cts/s, are excluded.
The source spectra are extracted from a circular region of a radius of $30^{\prime\prime}$ centered on the source,
whereas the background spectra are from a circular region of a radius of $45^{\prime\prime}$ in the same CCD chip near the source without chip edges or serendipitous sources, to minimize effects of non-real signals, such as Cu-K background emission lines.

In the analysis of the {\it Suzaku} data \citep{mit07}, we focus on the front-illuminated CCDs data of the X-ray Imaging Spectrometer (XIS0, XIS2, and XIS3; \citealt{koy07}).
which has a larger effective area around the energy band we focus on than the back-illuminated one (XIS1).
We reduced the data by using the {\tt HEASoft} v.6.23.
The data is screened with {\tt XSELECT} using the standard criterion.
The source spectra are extracted from circular regions of a radius of $2^{\prime}$  centered on the sources, whereas
the background spectra are from annuluses of $3^{\prime}-4^{\prime}$ in radii.
The response matrices and ancillary response files are generated for each XIS using {\tt xisrmfgen} and {\tt xisarfgen}.   

All the data are confirmed not to be affected by the pile-up effect or telemetry saturation.
The spectra are binned to have a minimum of 25 counts (50 counts for I Zw 1) per energy bin
to use the $\chi^2$ statistics in the spectral fitting.
The spectral fitting was carried out using {\tt xspec} v.12.10.0.

Parameters of the UFO lines are known to show time variability whose timescales are as fast as several days (e.g., \citealt{cap09,tom13}).
Because the equilibrium timescales of molecular outflows are much longer than days,
we need to calculate the {\it time-averaged} parameters of UFOs to compare them with those of molecular outflows.
On the other hand, some data have to be stacked in order to have enough sufficient photon statistics to be fitted.
Here, in the model fitting, we stack the X-ray data observed within two months, after checking that no significant changes of spectral features are seen.
We perform model fitting for each (stacked) energy spectrum, get the parameters, and calculate the time-averaged ones weighted by the exposure time.
We use 3.5--10.5~keV to focus on the Fe-K energy band and reduce the effect of neutral absorption (due to our Galaxy and AGN torus) and soft excess.
When the absorption is too strong, the {\tt ztbabs} model is added.
The spectral fitting process was carried out in a uniform way for all the observations.

\section{Results} \label{sec3}
In the spectral fitting of the X-ray archival data.
we started from a power-law continuum ({\tt pegpwrlw}) plus positive Gaussians to explain the emission lines, and record the chi-square ($\chi^2$).
In NGC 1068, which has the most complex reflection lines among the targets, 
we added six positive Gaussians at 6.4~keV (neutral Fe K$\alpha$), 6.7~keV (He-like Fe), 7.0~keV (H-like Fe), 7.4~keV (neutral Ni K$\alpha$), 7.8~keV (He-like Ni), and 8.2~keV (neutral Ni K$\beta$).
The upper panels of figure \ref{fig:spectra} show ratios of the data to the continuum models.
Then, we performed a blind search of absorption lines; we sequentially added a positive/negative Gaussian (with a fixed width of $\sigma=0.01$~keV) in the 4--10~keV band with a step of 0.1~keV, recorded the new chi-square ($\chi^2_{\rm n}$), and calculated $\Delta\chi^2=\chi^2_{\rm n}-\chi^2$.
The bottom panels of figure \ref{fig:spectra} show the $\Delta\chi^2$ plots.
The cyan, magenta, and blue levels are $\Delta\chi^2=-2.3$, $-4.61$, and $-9.21$,
which mean that the absorption lines are detected with 68\%, 90\%, and 99\% significance levels, respectively.
When the line significance of the negative Gaussian exceeds 90\%, we fitted the absorption lines with {\tt zxipcf} instead of the negative Gaussians, and derived the physical parameters of UFOs.
The {\tt zxipcf} model \citep{ree08} is a grid model for XSTAR photoionized absorption,
assuming a turbulent velocity of 200~km~s$^{-1}$.
This turbulent velocity is relatively lower than the typical UFOs in the Fe K band ($\sim1000$~km~s$^{-1}$; \citealt{tom11}), which may cause the absorption lines to saturate to high $N_{\rm H}$. 
We fixed the covering fraction of this model as unity, because the covering fraction and the column density degenerate in the optically-thin case \citep{miz14,miz17}.
The red lines in figure \ref{fig:spectra} show the final model.
Consequently, the UFO lines are detected in six targets.
Their properties are listed in table \ref{tab:prop}.
Table \ref{tab:mol_outflow} shows the parameters of the molecular outflows in literature \citep{cic12}.
The molecular outflows in NGC 1068 and NGC 6240 are spatially resolved by Atacama Large Millimeter/submillimeter Array (ALMA), and their published results \citep{gar14,sai18} are also listed.
In addition to this, the results of IRAS F11119+3257 are shown.

The BH masses and bolometric luminosities are needed to calculate the Eddington ratio (table \ref{tab:prop}). Several methods are used to constrain them.
One of the most accurate methods is to make a rotation curve of the maser emission around the central BH, which determined the mass of NGC 1068  \citep{gre97}.
The masses of I Zw 1 and IRAS F11119+3257 are estimated from the virial theorem, in which the full-width half-maximum of H$\beta$ broad-line emission and the radius of the broad-line region estimated by $L_\lambda(5100\AA)$ \citep{ves02,kaw07}.
The mass of IC 5063 is estimated from the relation to the bulge luminosity \citep{nic03}, and
those of Mrk 231, Mrk 273 and NGC 6240 are from the one to the bulge dispersion \citep{das06}.
We use the intrinsic bolometric luminosities listed in \citet{cic14} (and \citealt{tom15} for IRAS F11119+3257).

\begin{deluxetable*}{lcccccc}
\tablewidth{0pt} 
\tablecaption{Properties of AGNs \label{tab:prop}}
\tablehead{
\colhead{Object}&
\colhead{Type}&
\colhead{$z$}&
\colhead{$M_{\rm BH}$}&
\colhead{$\log(L_{\rm AGN})$}&
\colhead{$L_{\rm AGN} / L_{\rm Edd}$}&
\colhead{Refs.}\\
\colhead{}&
\colhead{}&
\colhead{}&
\colhead{($M_\odot$)}&
\colhead{(erg s$^{-1}$)}&
\colhead{}&
\colhead{}}
\startdata
IC 5063  & Seyfert 2 & 0.0110 & $2.8\times10^8$ & 44.3 & $5.7\times10^{-3}$ &1, 2\\
I Zw 1 & NLSy 1 & 0.0611 & $1.8\times10^7$ & 45.4 & $1.1$ &1, 3\\
Mrk 231  & Seyfert 1 & 0.0422 & $1.7\times10^7$ & 45.7 & 2.5  &1, 4\\
Mrk 273  & Seyfert 2 & 0.0378 & $5.6\times10^8$ & 44.7 & $7.6\times10^{-3}$ &1, 4\\
NGC 1068 & Seyfert 2 & 0.0038 & $1.5\times10^7$ & 43.9 & $4.2\times10^{-2}$ &1, 5\\
NGC 6240 & Seyfert 2 & 0.0245 & $2.3\times10^8$ & 45.4 & $8.3\times10^{-2}$ &1, 4\\
\hline
IRAS F11119+3257&Type 1 quasar &0.189&$1.6\times10^7$&46.2&7.9&6, 7\\
\enddata
\tablecomments{We use the BH masses and AGN luminosities in literature, and calculate Eddington ratios from these values.}
\tablerefs{(1) \citet{cic14} and references therein; (2) \citet{nic03}; (3) \citet{ves02}; (4) \citet{das06}; (5) \citet{gre97}; (6) \citet{kaw07}; (7) \citet{tom15} }
\end{deluxetable*}

\begin{deluxetable*}{lccccc}
\tablewidth{0pt} 
\tablecaption{Properties of the molecular outflows \label{tab:mol_outflow}}
\tablehead{
\colhead{Object}&
\colhead{$v_{\rm mol}$} & 
\colhead{$\dot{M}_{\rm mol}$} & 
\colhead{$\dot{P_{\rm mol}}$} & 
\colhead{$\dot{K_{\rm mol}}$} & 
\colhead{Refs.}\\
\colhead{}&
\colhead{(km s$^{-1}$)} & 
\colhead{($M_\odot$~yr$^{-1}$)} & 
\colhead{$(L_{\rm AGN}/c)$} & 
\colhead{($L_{\rm AGN}$)}&
\colhead{}}
\startdata
IC 5063  &300 &23--127 &7--36 &$[4-18]\times10^{-3}$&1\\
I Zw 1 &(500)&$\leq140$&$\leq6$&$\leq5\times10^{-3}$&1\\
Mrk 231 &700 &1050  &26  &$3\times10^{-2}$&1\\
  Mrk 273  &620 &600   &130 &0.1&1\\
NGC 1068 &150 &84   &27 &$2\times10^{-2}$&1\\
\,\,\,\,(ALMA) &75 & $60^{+20}_{-40}$   &$23^{+8}_{-14}$ &$6^{+2}_{-4}\times10^{-3}$&2\\
NGC 6240  &400 &800   &25  &$2\times10^{-2}$&1\\
\,\,\,\,(ALMA) &500 & $230$   & 8&$7\times10^{-3}$&3\\
\hline
IRAS F11119+3257&1000&80--200&1.5--3.0&$1.5-4.0\times10^{-3}$&4\\
\enddata
\tablerefs{(1) \citet{cic14} and references therein; (2) \citet{gar14}; (3) \citet{sai18}; (4) \citet{vei17}}
\end{deluxetable*}

\begin{figure*}
\centering
\subfigure{
\resizebox{6cm}{!}{\includegraphics[angle=270]{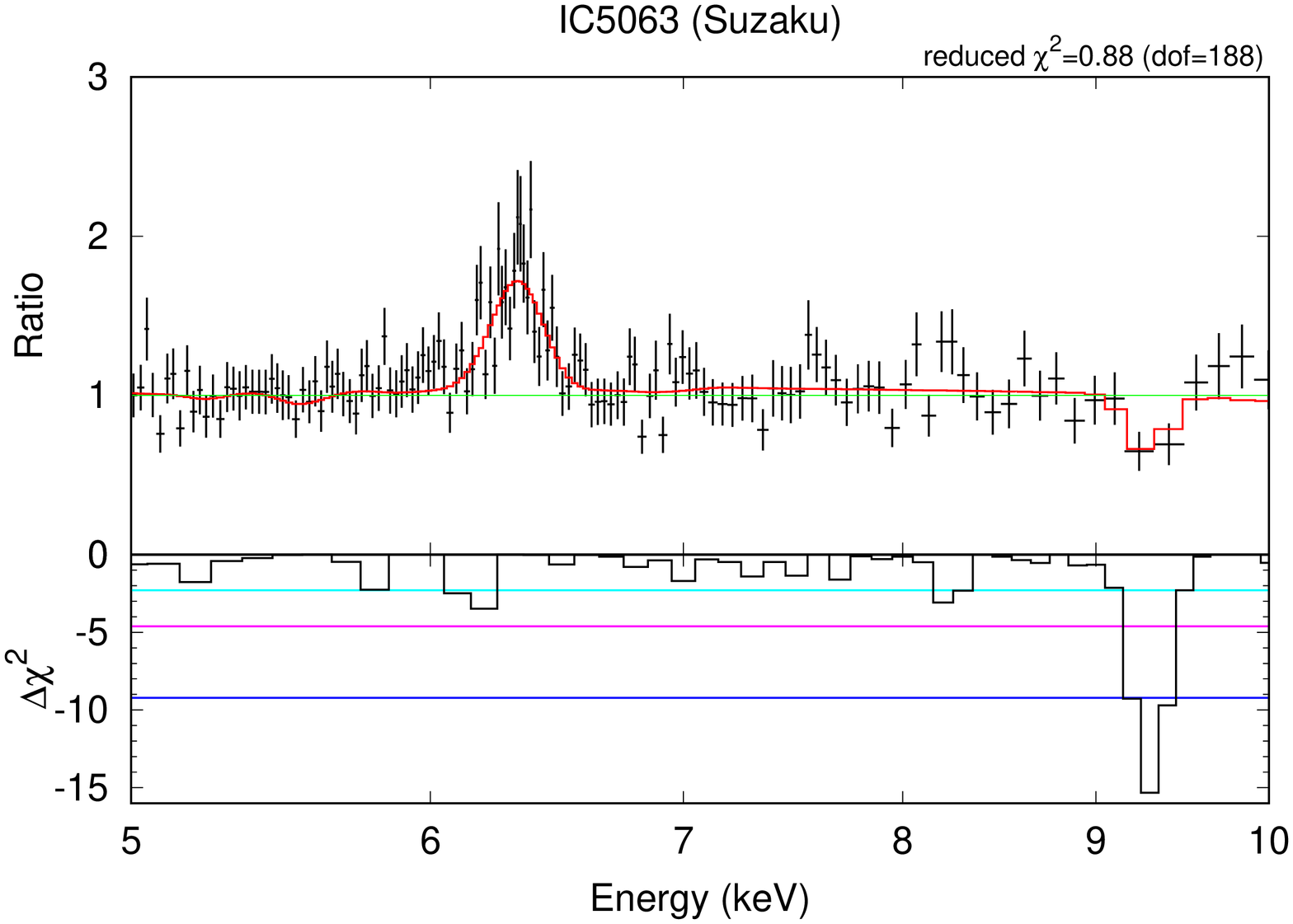}}
\resizebox{6cm}{!}{\includegraphics[angle=270]{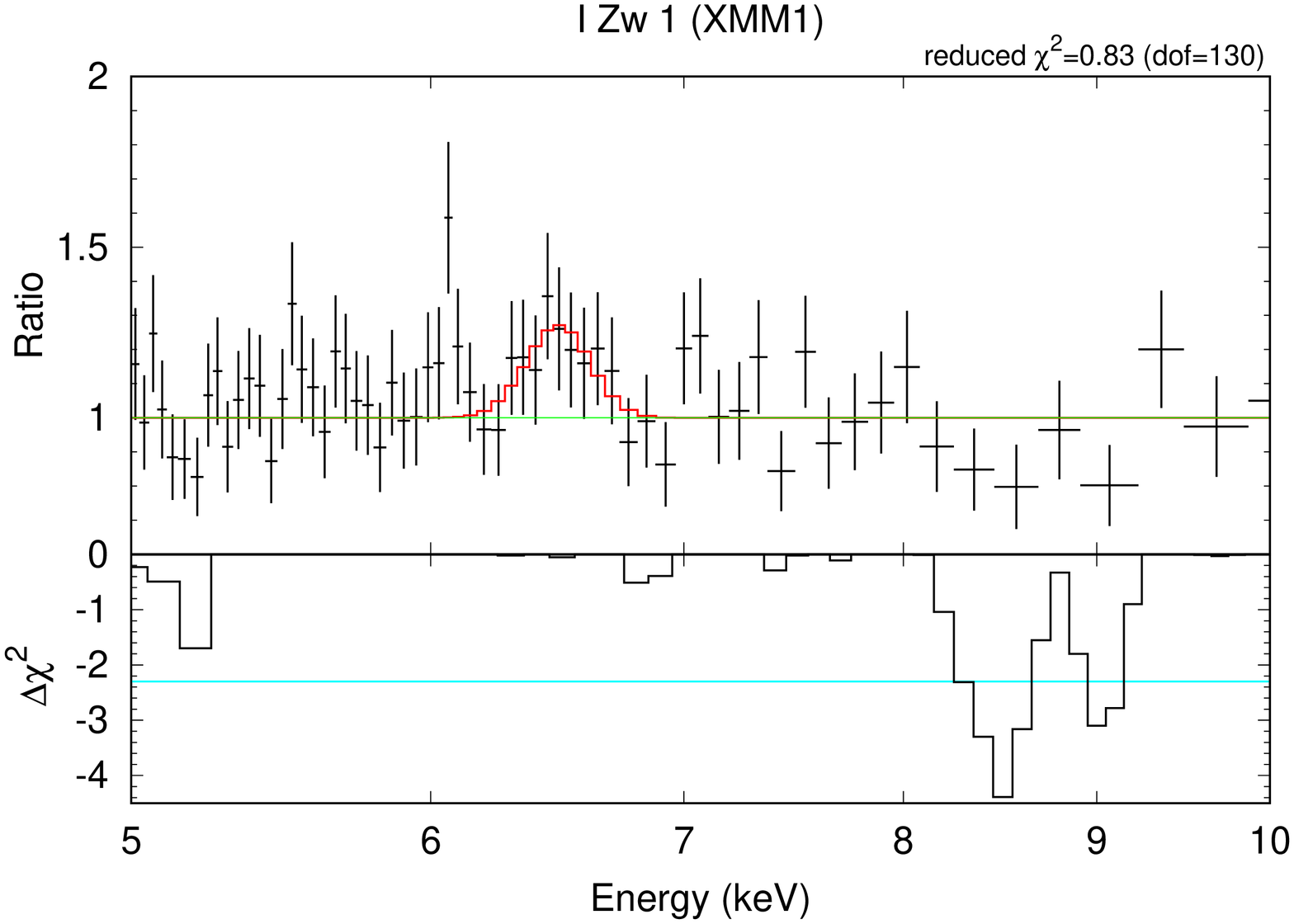}}
\resizebox{6cm}{!}{\includegraphics[angle=270]{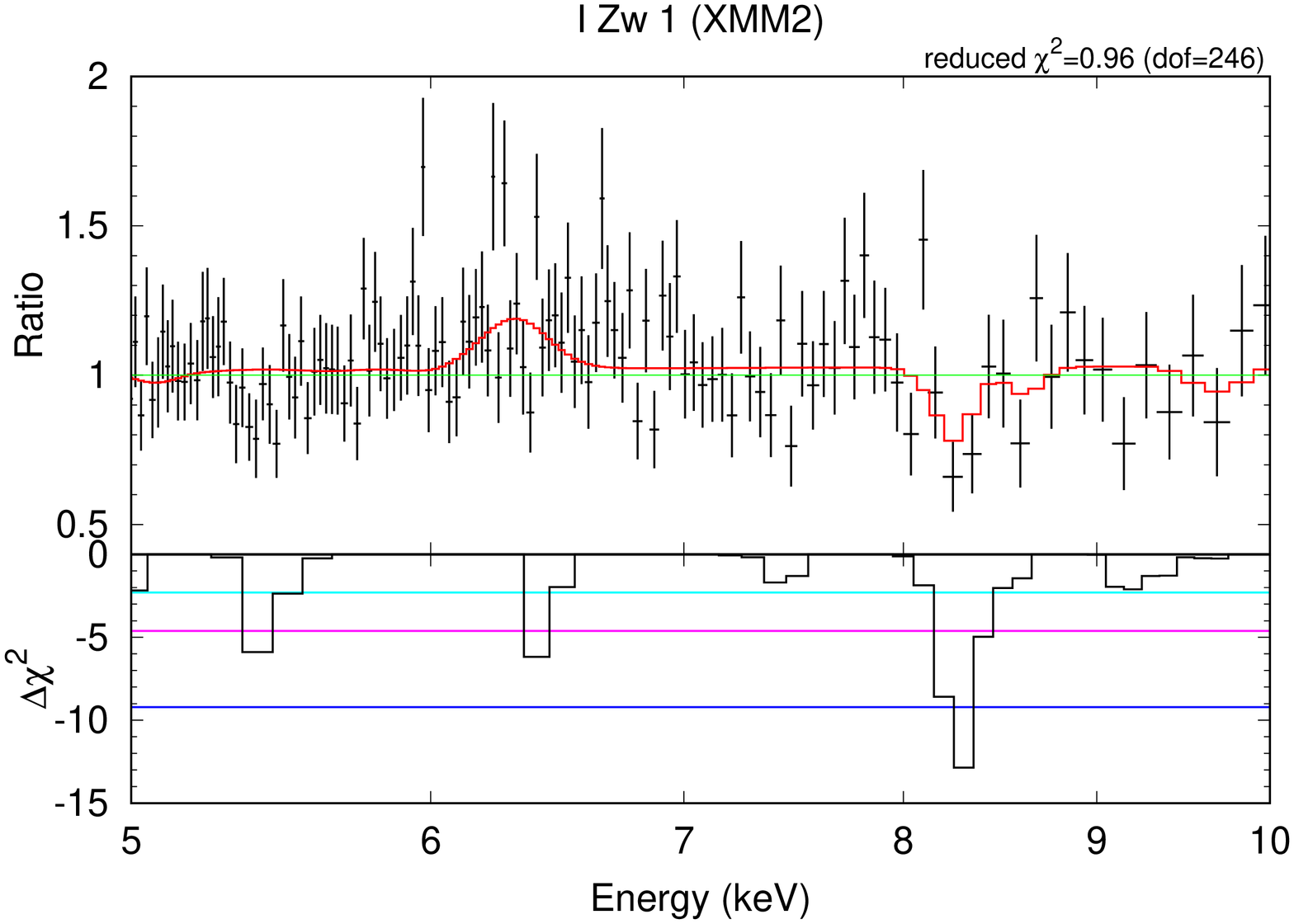}}
}
\subfigure{
\resizebox{6cm}{!}{\includegraphics[angle=270]{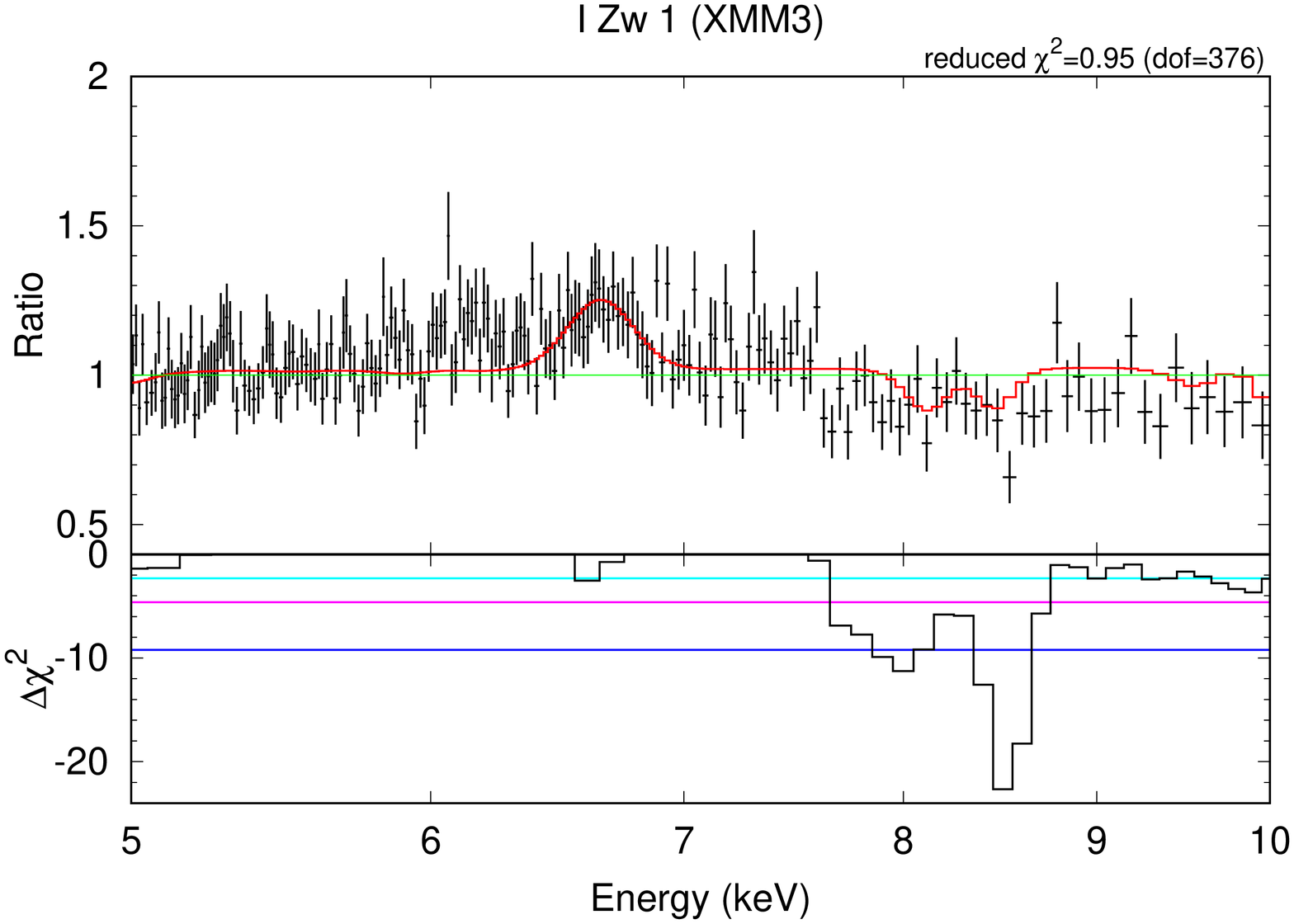}}
\resizebox{6cm}{!}{\includegraphics[angle=270]{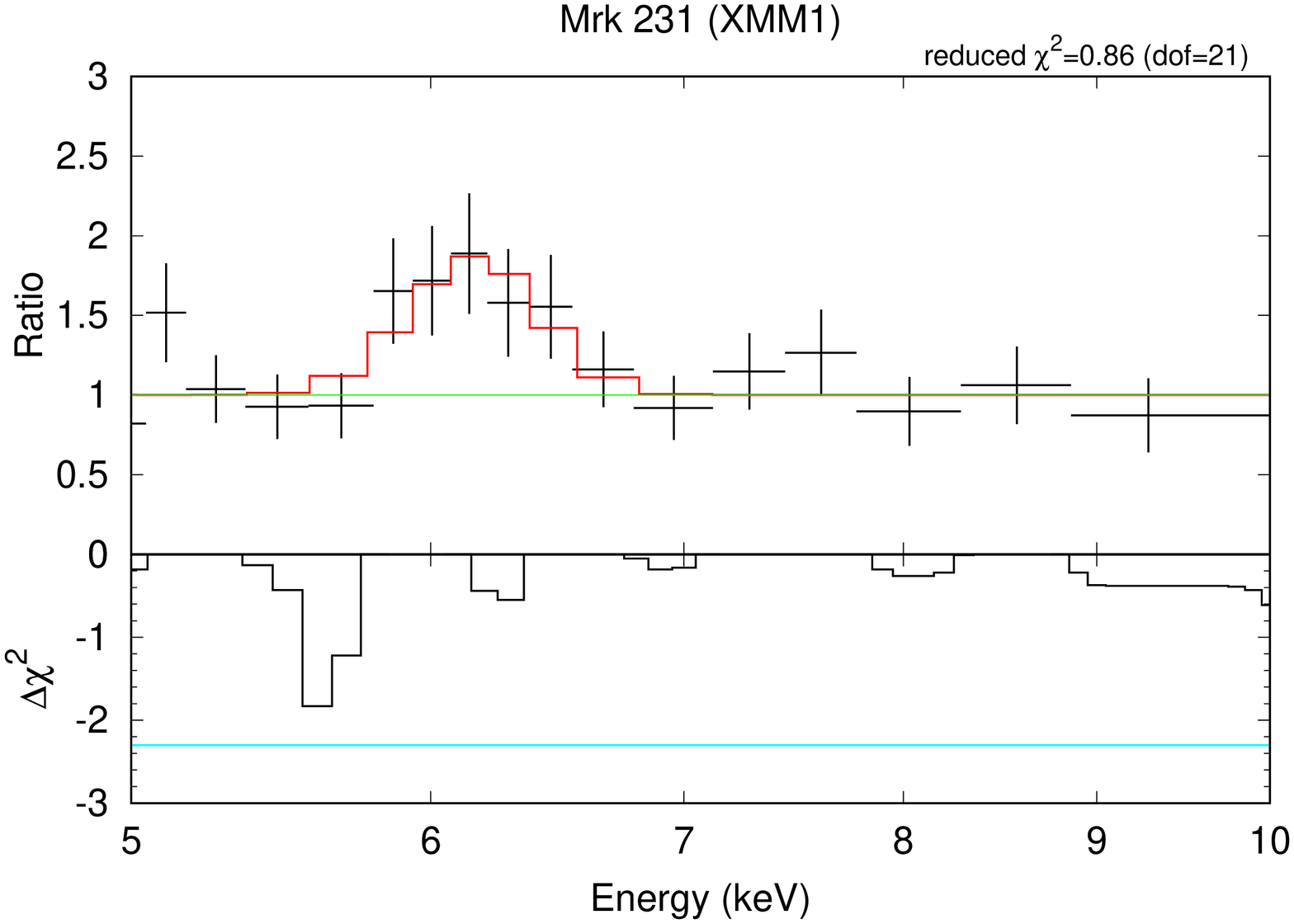}}
\resizebox{6cm}{!}{\includegraphics[angle=270]{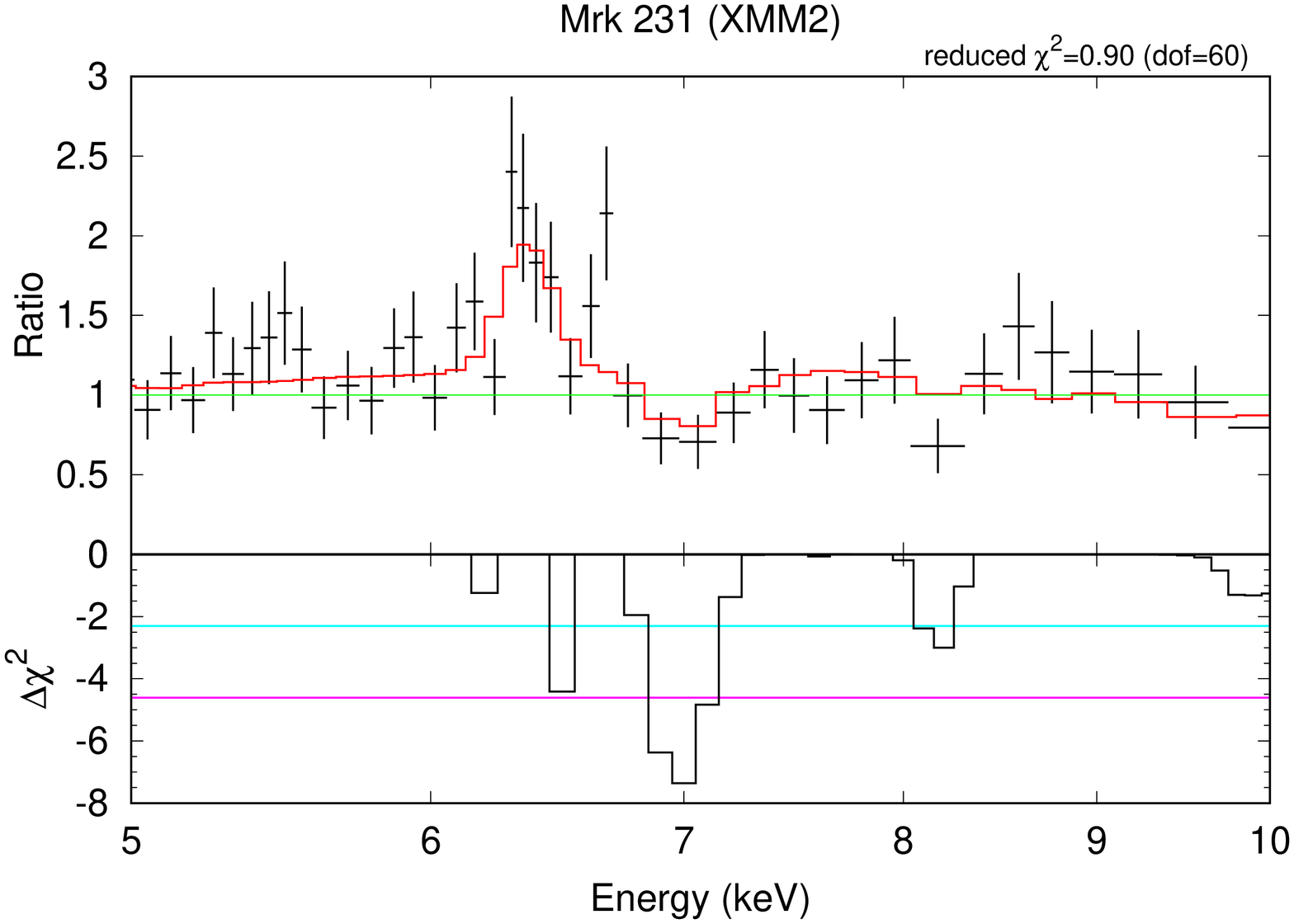}}
}
\subfigure{
\resizebox{6cm}{!}{\includegraphics[angle=270]{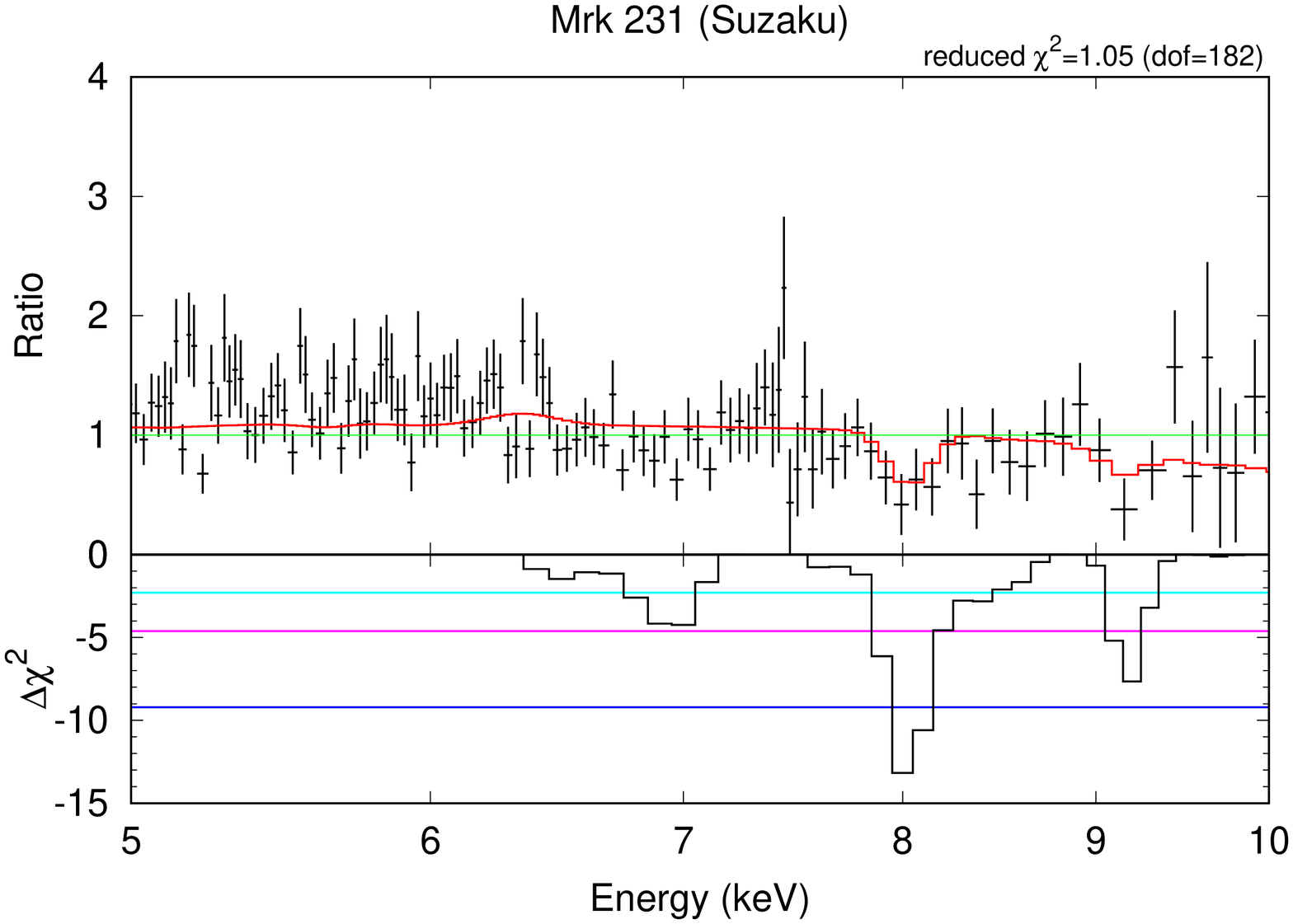}}
\resizebox{6cm}{!}{\includegraphics[angle=270]{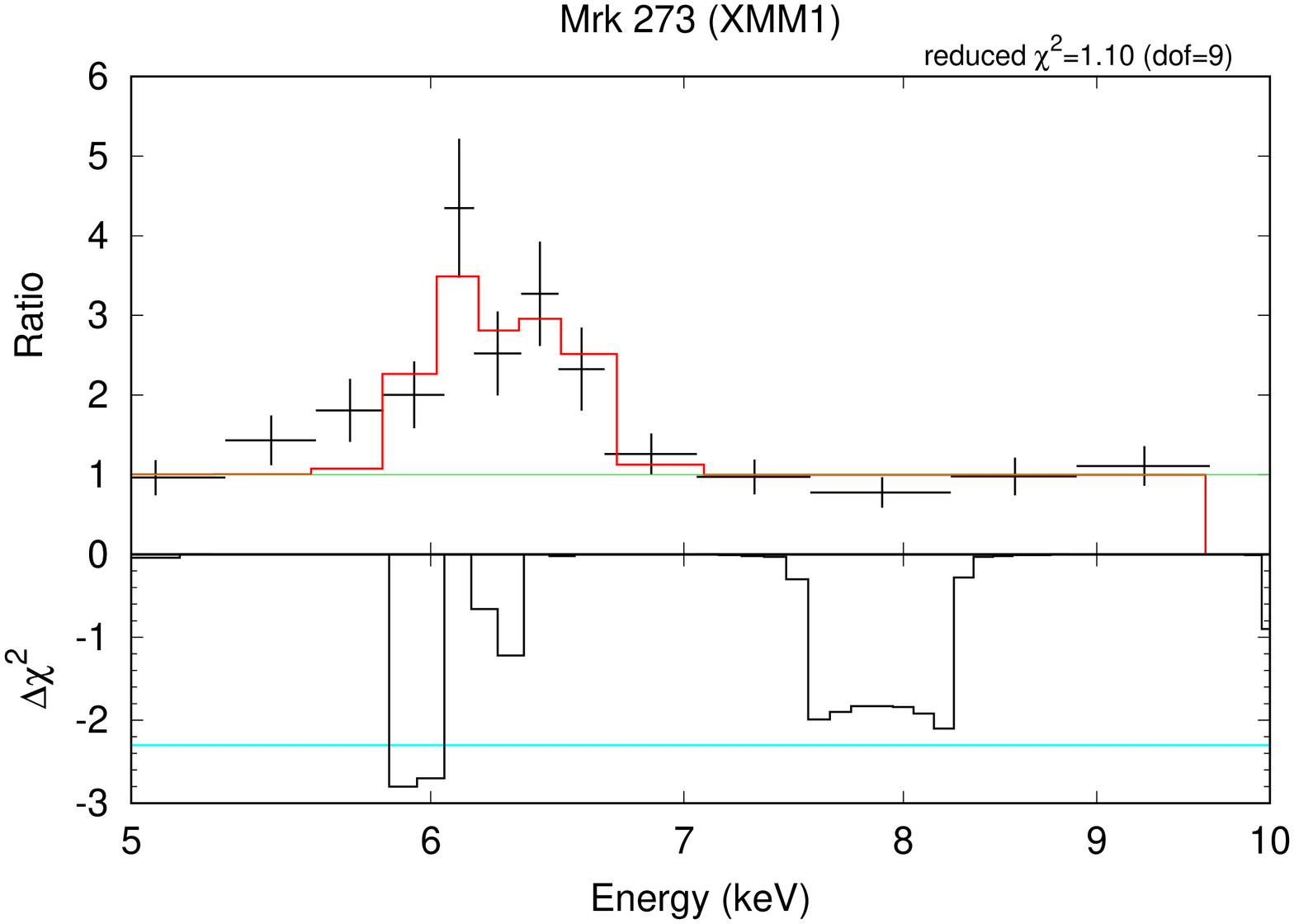}}
\resizebox{6cm}{!}{\includegraphics[angle=270]{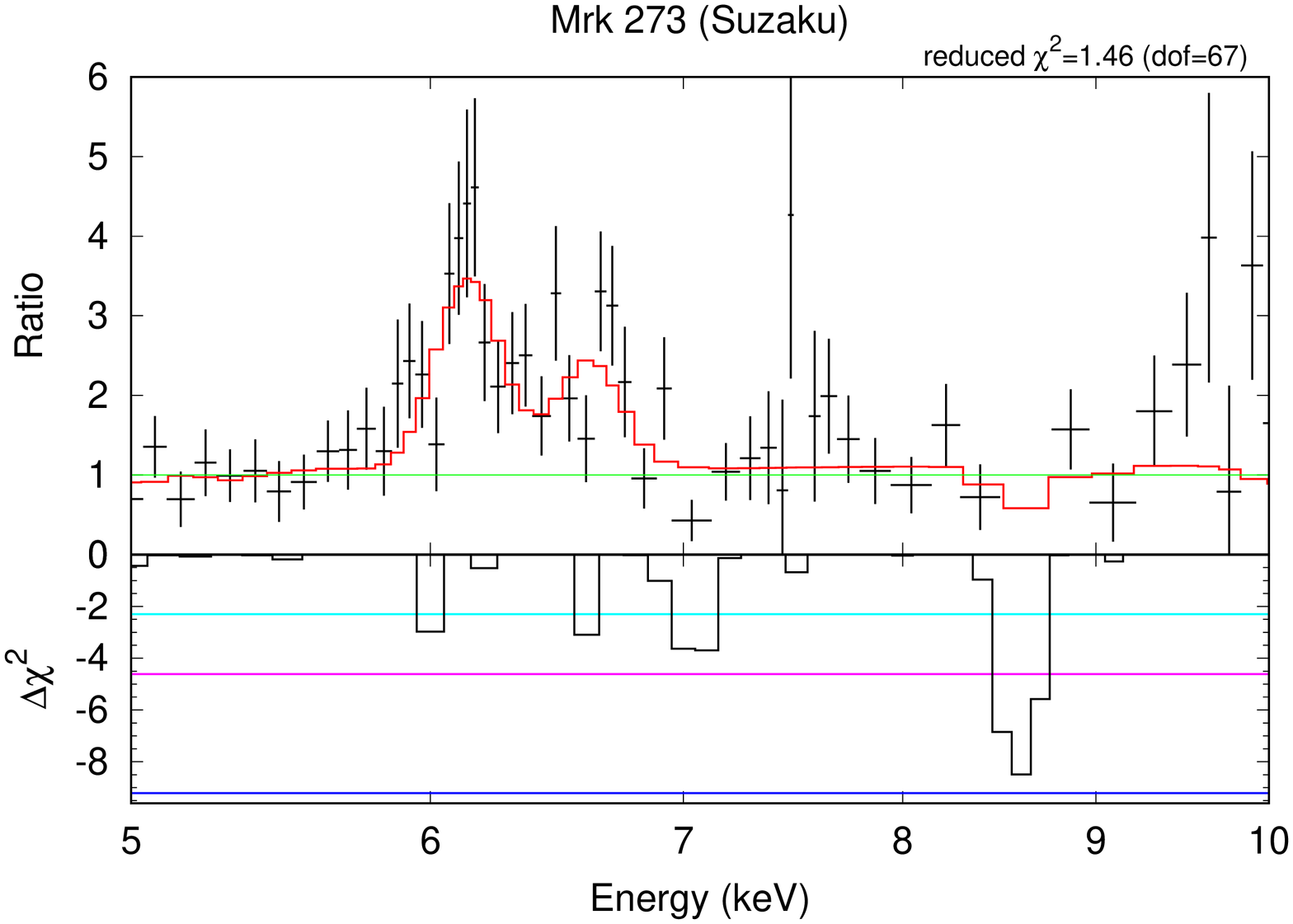}}
}
\subfigure{
\resizebox{6cm}{!}{\includegraphics[angle=270]{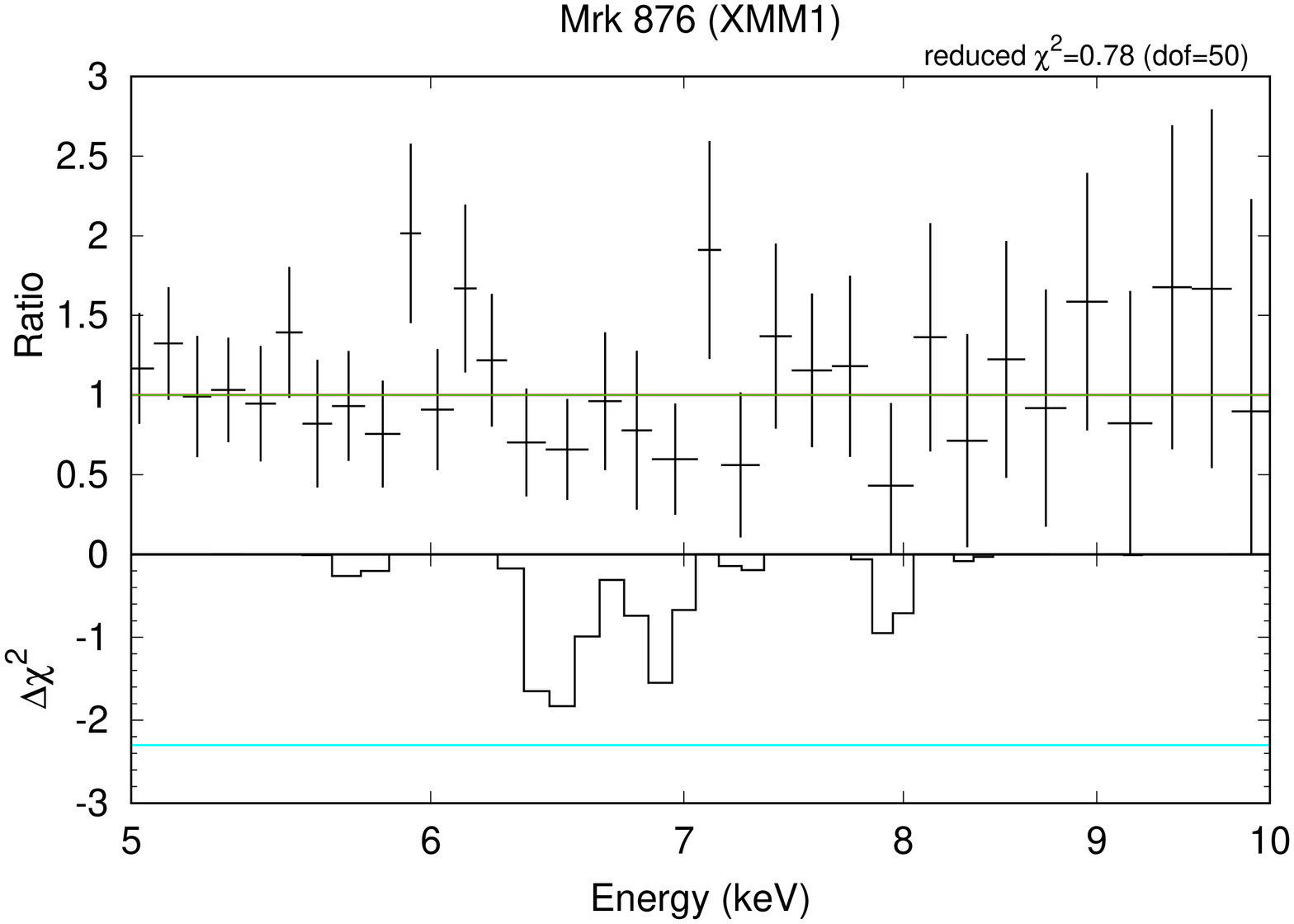}}
\resizebox{6cm}{!}{\includegraphics[angle=270]{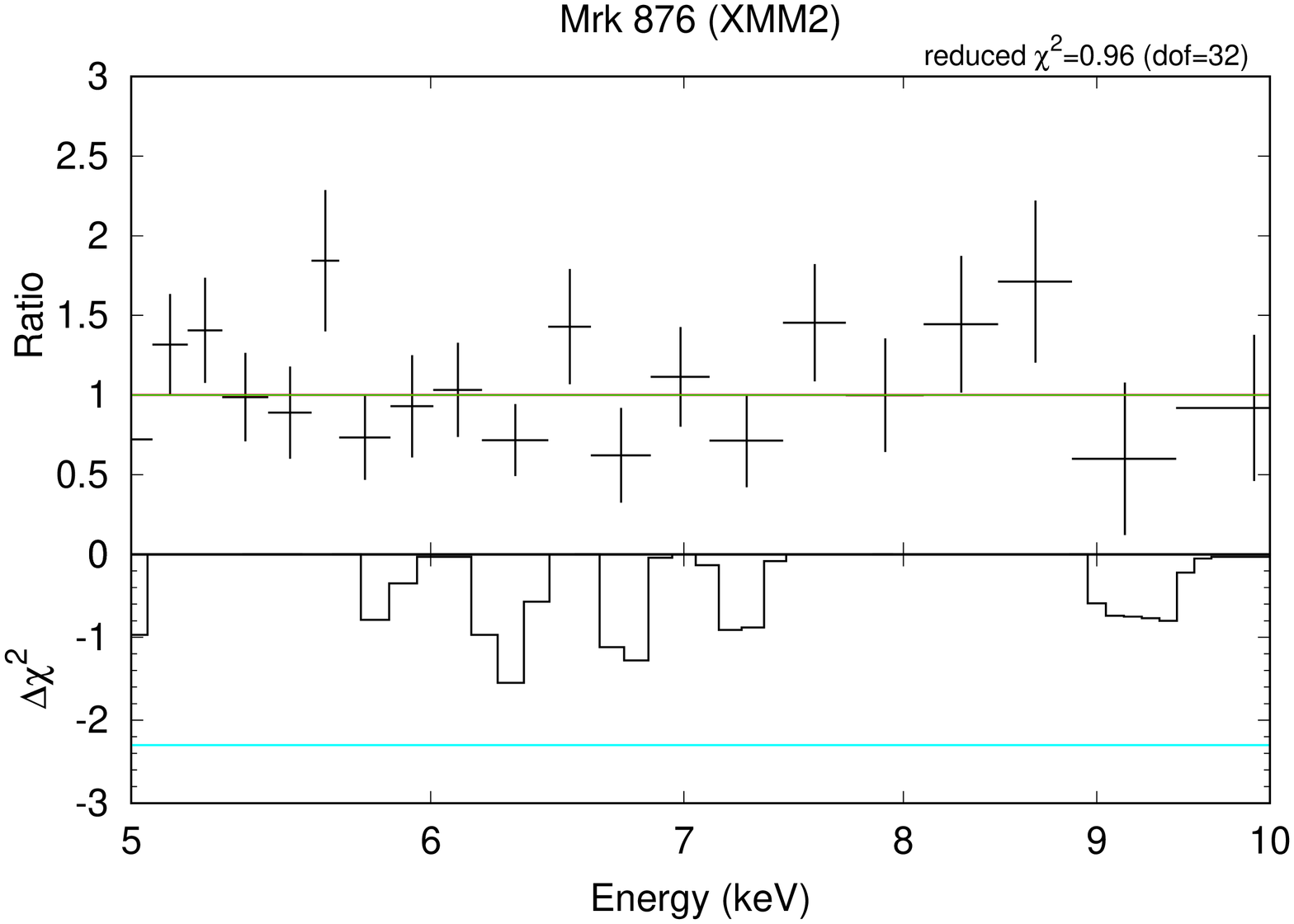}}
\resizebox{6cm}{!}{\includegraphics[angle=270]{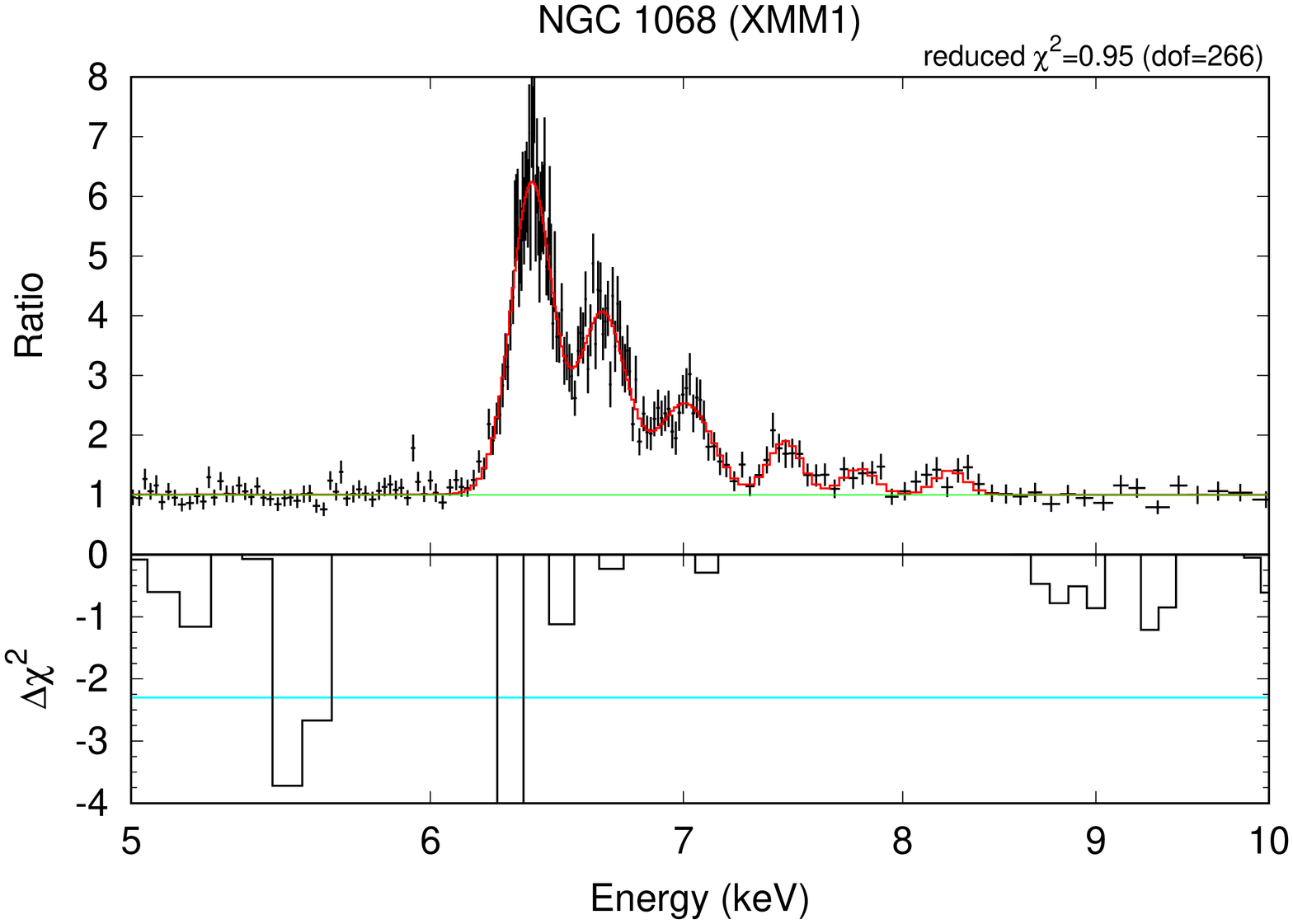}}
}
\subfigure{
\resizebox{6cm}{!}{\includegraphics[angle=270]{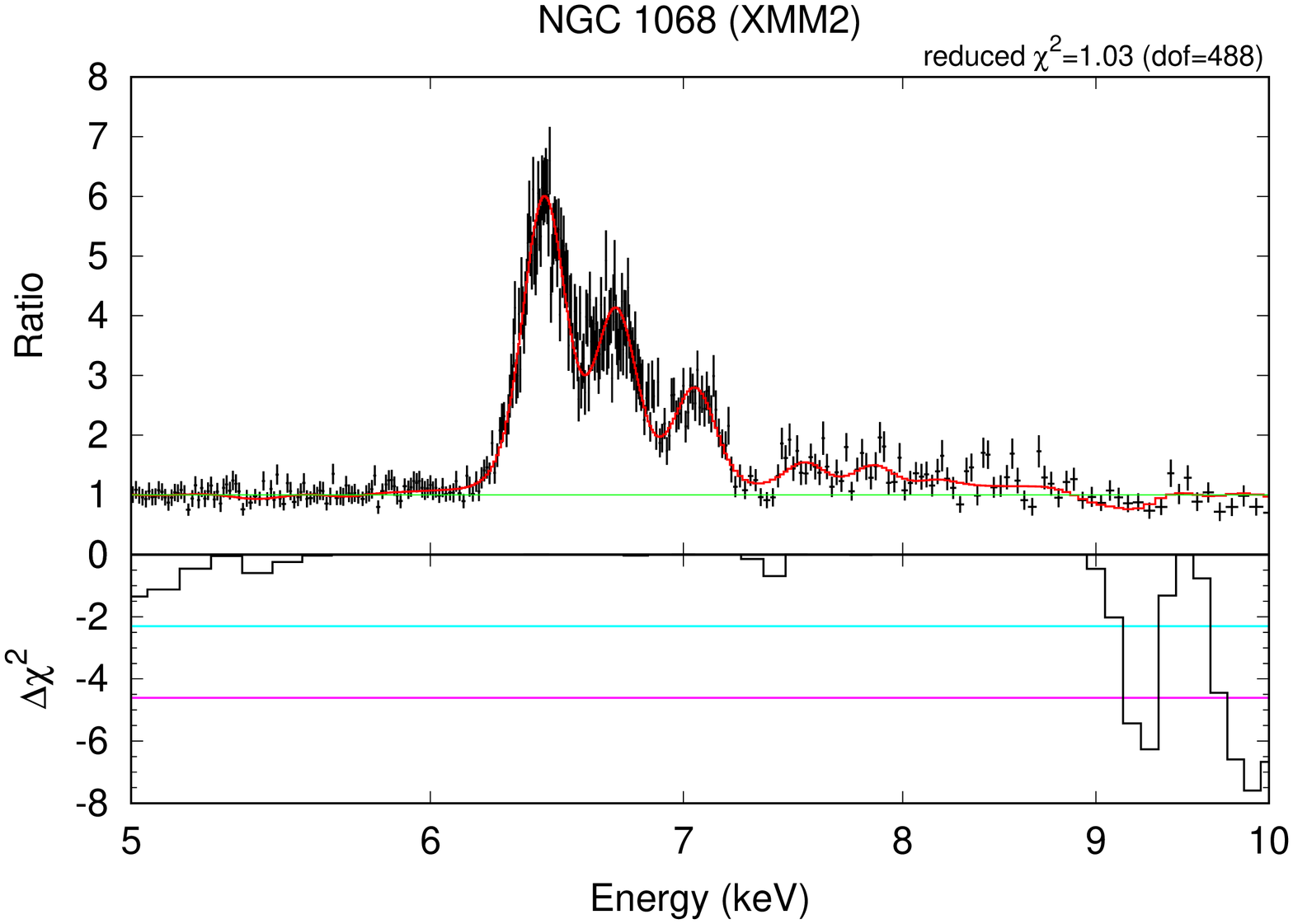}}
\resizebox{6cm}{!}{\includegraphics[angle=270]{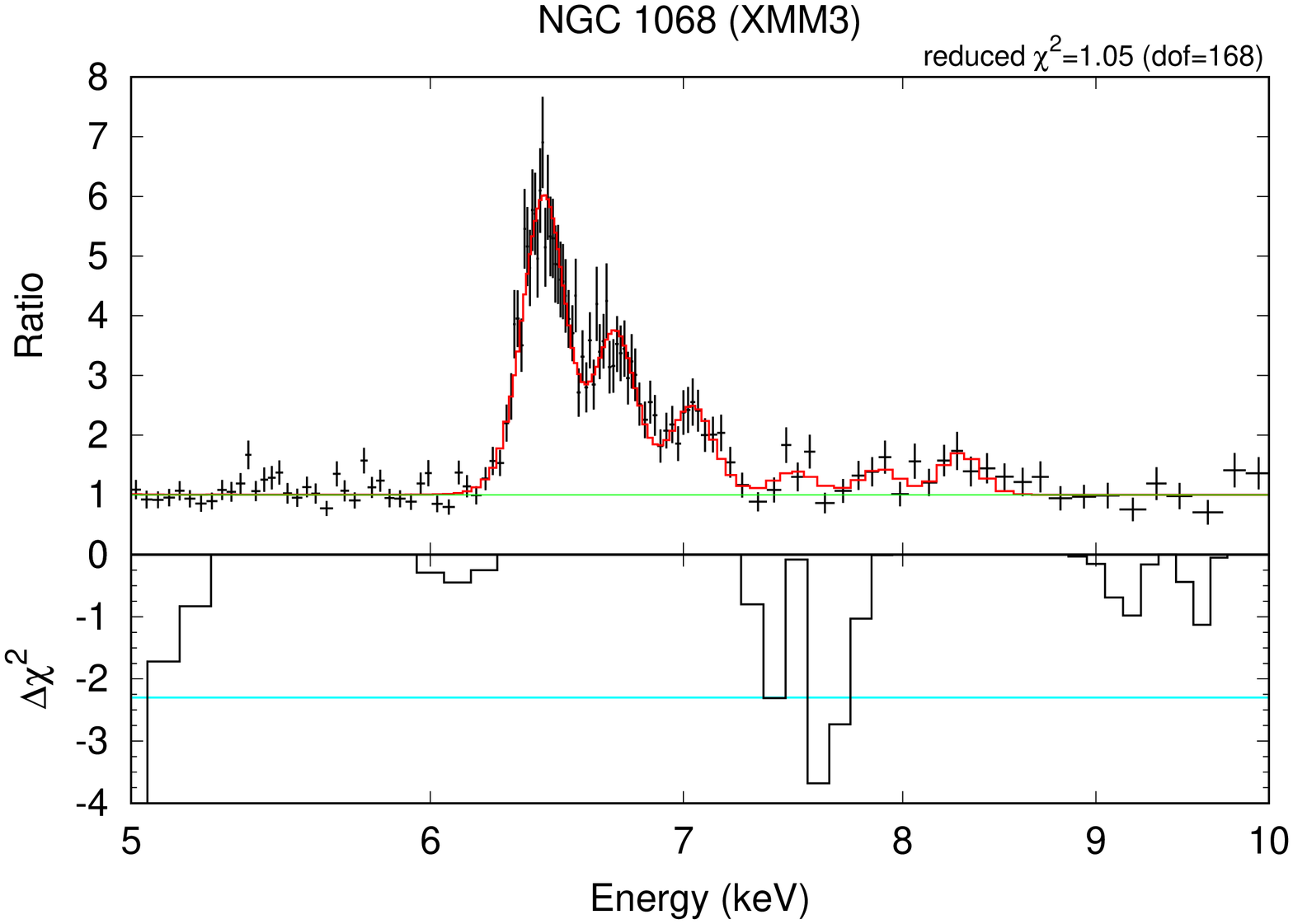}}
\resizebox{6cm}{!}{\includegraphics[angle=270]{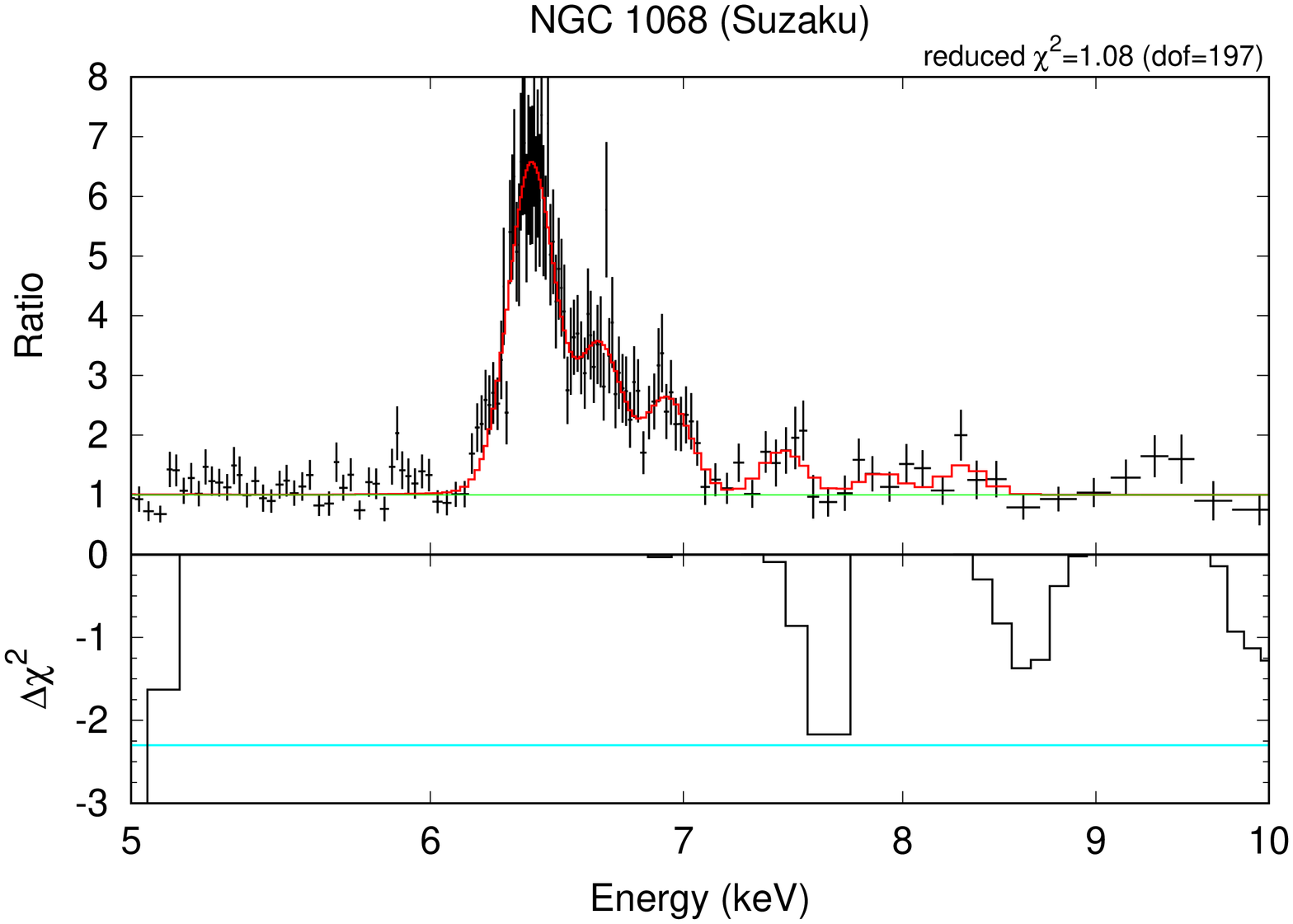}}
}
\caption{Top: ratio of the energy spectra to the continuum (power-law) component.
The red curvatures show the fitting model.
The absorption features can be seen in the red lines when the absorption lines are detected more than 90\% significance level.
The reduced $\chi^2$ and degrees of freedom are shown in the top right of each panel.
Bottom: $\Delta \chi^2$ plots with the 68\% (cyan), 90\% (magenta), and 99\% (blue) significance levels, from top to bottom. Only the values when the normalization of the Gaussian is negative are plotted.}
\label{fig:spectra}
\end{figure*}

\addtocounter{figure}{-1}
\begin{figure*}
\addtocounter{subfigure}{1}
\centering
\subfigure{
\resizebox{6cm}{!}{\includegraphics[angle=270]{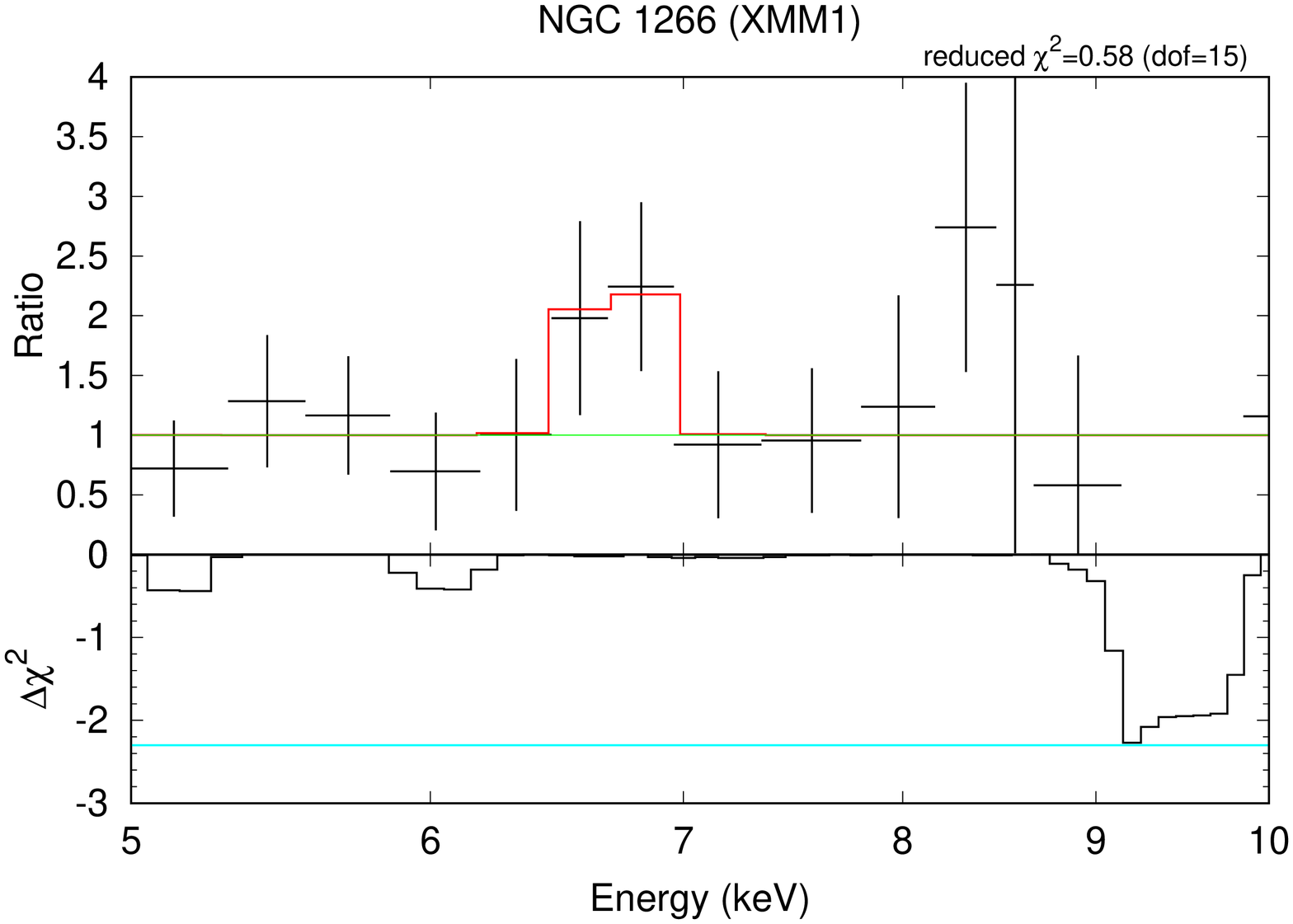}}
\resizebox{6cm}{!}{\includegraphics[angle=270]{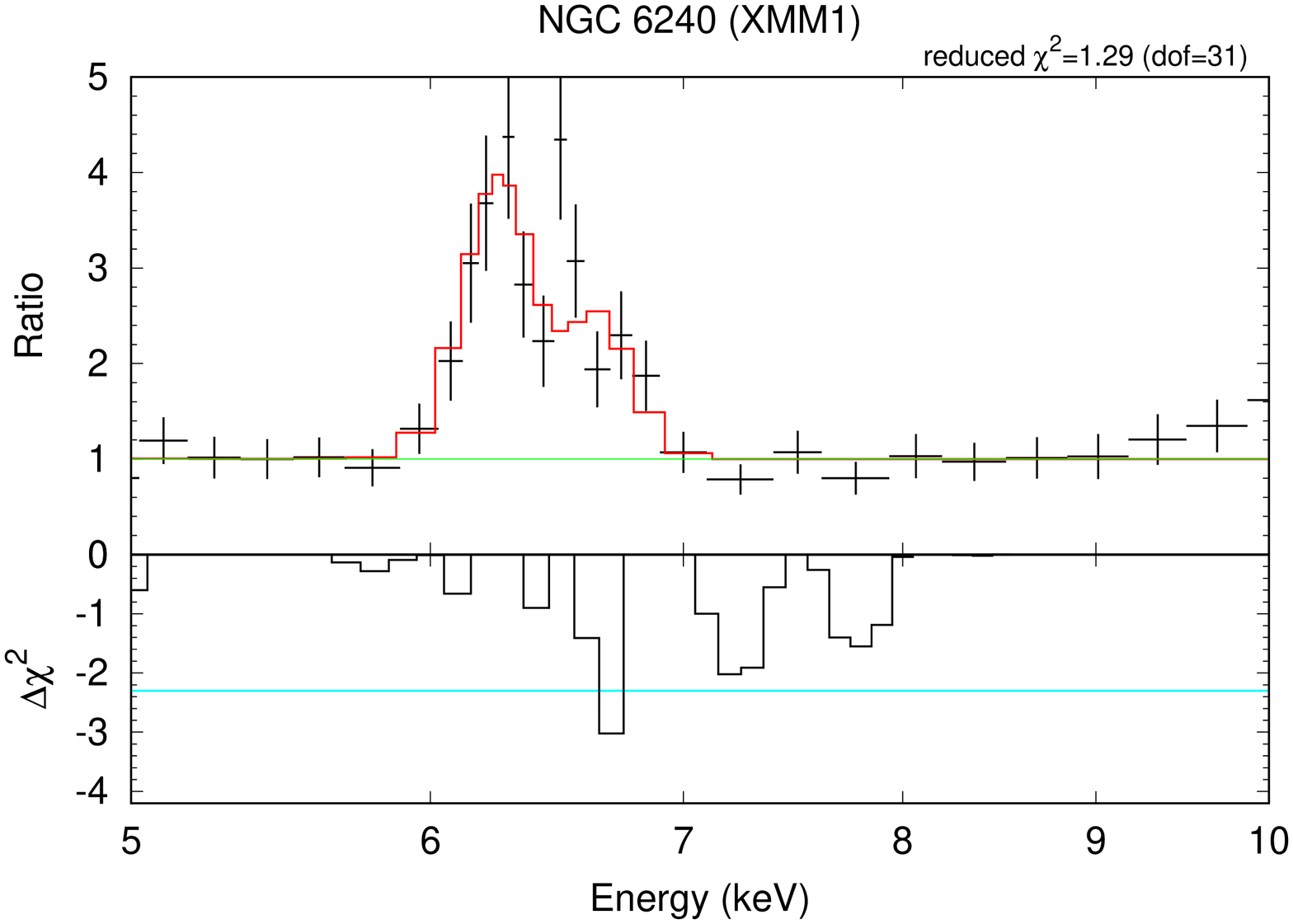}}
\resizebox{6cm}{!}{\includegraphics[angle=270]{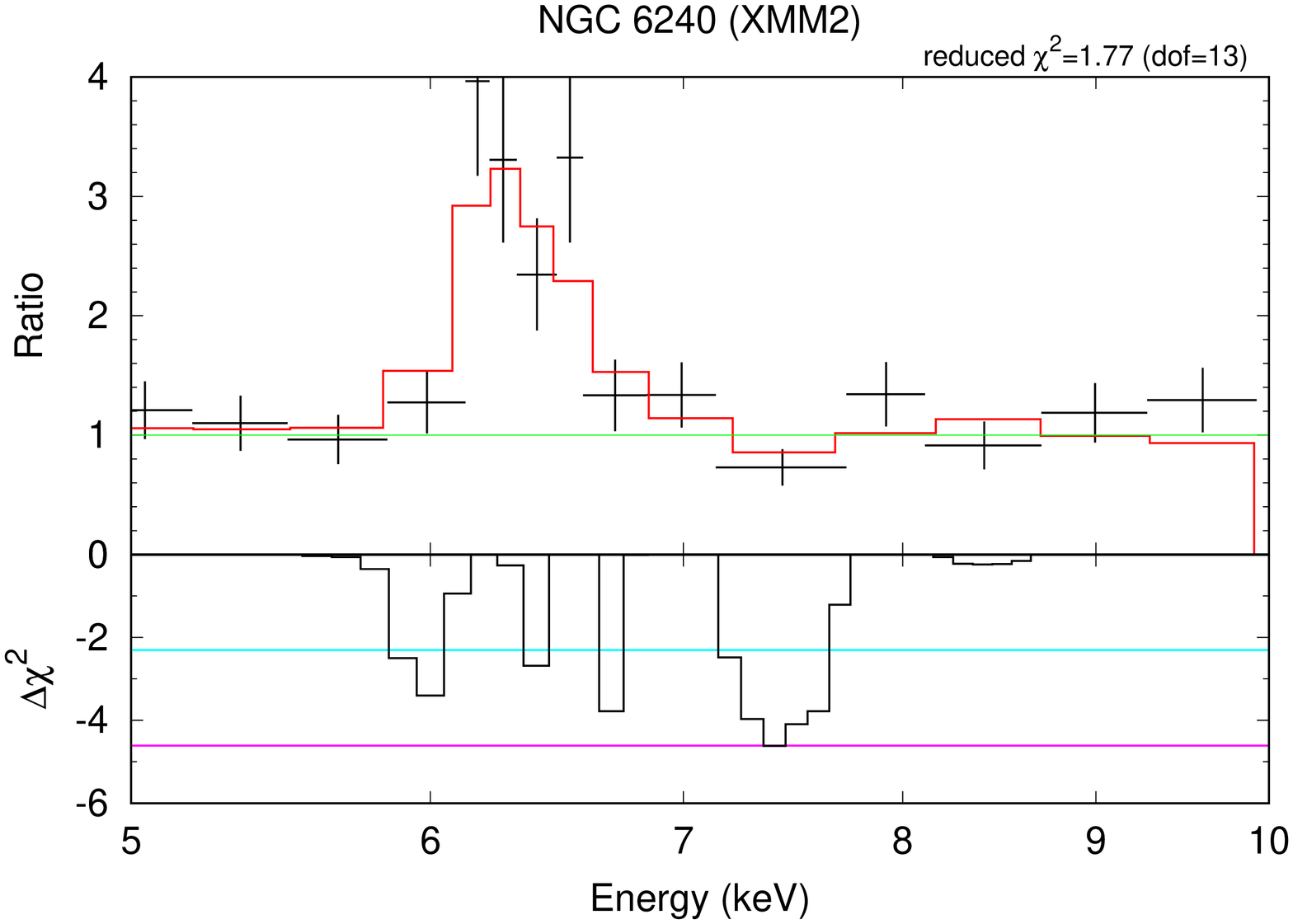}}
}
\subfigure{
\resizebox{6cm}{!}{\includegraphics[angle=270]{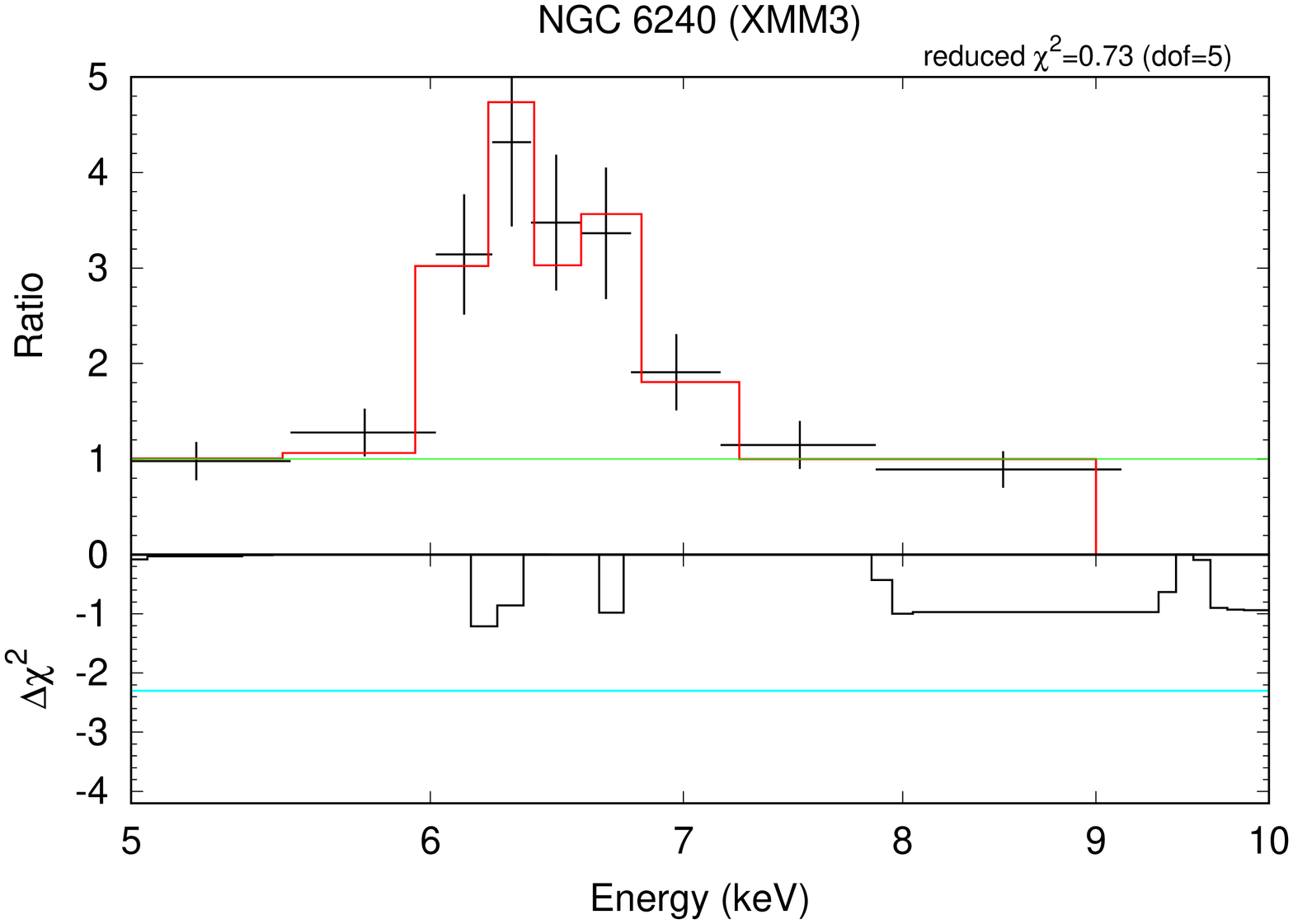}}
\resizebox{6cm}{!}{\includegraphics[angle=270]{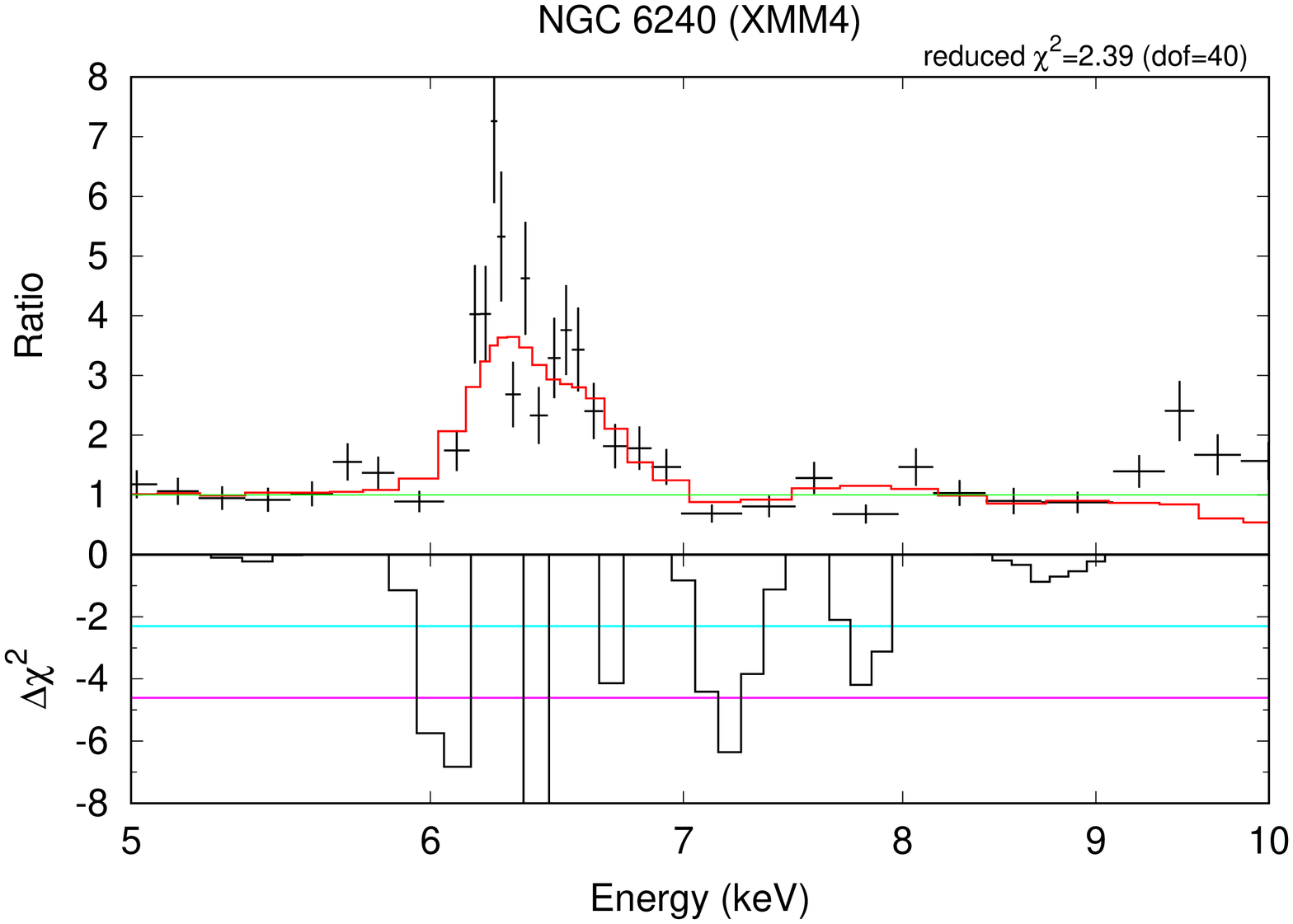}}

}
\caption{{\it Continued.}}
\label{fig:spectra2}
\end{figure*}

The mass outflow rate, momentum, and energy-loss rate of the UFO are expressed as
\begin{equation}
\begin{split}
\dot{M}_{\rm UFO}&\sim\Omega b r^2 m_p n(r)v_{\rm UFO}\\
\dot{P}_{\rm UFO}&\sim\dot{M}_{\rm UFO}v_{\rm UFO}\\
\dot{K}_{\rm UFO}&\sim\dot{M}_{\rm UFO}v^2_{\rm UFO}/2,
\end{split}\label{eq1}
\end{equation} 
where $\Omega$ is the solid angle of the wind, $b$ the volume filling factor, $m_p$ is the proton mass, and $n(r)$ is the electron number density \citep{gof15}.
They explicitly depend on $r$, which is difficult to directly constrain with the observables.
In this paper, we use a conservative manner that the wind is launched at the radius where the wind velocity exceeds the escape velocity (e.g. \citealt{tom15}).
When $v_{\rm UFO}$ equals the escape velocity $v_{\rm esc}=\sqrt{2GM_{\rm BH}/r}$,
the location of the wind is written as
\begin{equation}
r=\frac{2GM_{\rm BH}}{v_{\rm UFO}^2}.
\end{equation}
Please note that this radius gives a minimum value of equation (\ref{eq1}).
In this case, equation (\ref{eq1}) are 
\begin{equation}
\begin{split}
\dot{M}_{\rm UFO}&\sim2\Omega G M_{\rm BH}m_pN_{\rm H}v_{\rm UFO}^{-1}\\
\dot{P}_{\rm UFO}&\sim2\Omega G M_{\rm BH}m_pN_{\rm H}\\
\dot{K}_{\rm UFO}&\sim\Omega G M_{\rm BH}m_pN_{\rm H} v_{\rm UFO},
\end{split}\label{eq2}
\end{equation}
respectively, (see equation 3--5 in \citealt{gof15}).
We adopt $\Omega/4\pi=0.4$ as the typical and conservative value, because the detection rate of the UFO lines in the literature is about 40\% \citep{tom10,gof13}.
The derived values in each observation are listed in table \ref{tab:UFO}.

\begin{deluxetable*}{llcccccc}[b!]
\tablecaption{Results of the X-ray spectral fitting of UFO lines in each data set \label{tab:UFO}}
\tablecolumns{15}
\tablewidth{0pt}
\tablehead{
\colhead{Object}&\colhead{Name}&\colhead{$N_{\rm H}$}&\colhead{$\log\xi$}&\colhead{$v_{\rm UFO}$}&\colhead{$\dot{M}_{\rm UFO}$}&\colhead{$\dot{P}_{\rm UFO}$}&\colhead{$\dot{K}_{\rm UFO}$}\\
\colhead{}&\colhead{}&\colhead{(cm$^{-2}$)}&\colhead{}&\colhead{(km s$^{-1}$)}&\colhead{($M_\odot$~yr$^{-1}$)}&\colhead{($L_{\rm AGN}/c$)}&\colhead{($L_{\rm AGN}$)}
}
\startdata
IC 5063  & Suzaku & $1.2_{-0.6}^{+1.2}\times10^{23}$ & $2.7_{-0.3}^{+0.2}$  &$9.29_{-0.14}^{+0.13}\times10^4$
&  $1.3_{-0.6}^{+1.3}\times10^{-1}$&$1.2_{-0.5}^{+1.1}\times10^1$&$1.8_{-0.8}^{+1.7}$\\
I Zw 1   & XMM2 & $5.0_{-4.3}^{+13.9}\times10^{22}$  & $3.6_{-1.0}^{+0.7}$    & $8\pm2\times10^4$
&$6.5_{-5.6}^{+18.0}\times10^{-3}$&$4.0_{-3.4}^{+10.9}\times10^{-2}$&$5.0_{-4.3}^{+13.7}\times10^{-3}$\\
& XMM3   & $2.3_{-1.9}\times10^{23}$  & $4.4_{-0.4}$  & $7.1\pm0.3\times10^4$
&$3.2_{-2.7}\times10^{-2}$&$1.8_{-1.5}\times10^{-1}$&$2.2_{-1.8}\times10^{-2}$\\
Mrk 231  & XMM2  &  $8_{-5}^{+12}\times10^{23}$  & $3.4_{-0.3}^{+0.5}$  & $12.7_{-0.4}^{+1.3}\times10^4$
&$1.7_{-1.4}^{+2.8}\times10^{-1}$&$1.6_{-1.1}^{+2.7}\times10^{-1}$&$7_{-5}^{+13}\times10^{-3}$\\
 & Suzaku  &  $8_{-6}^{+22}\times10^{22}$  & $2.7\pm0.9$      & $7.0\pm0.3\times10^4$
 &$7_{-5}^{+19}\times10^{-3}$&$1.6_{-1.4}^{+4.8}\times10^{-2}$&$1.9_{-1.6}^{+5.6}\times10^{-3}$\\
Mrk 273 & Suzaku  &  $1.8_{-1.6}^{+3.1}\times10^{24}$  & $3.5_{-0.6}^{+0.9}$ &$7.9\pm0.3\times10^4$
&$4.5_{-3.9}^{+7.8}$&$1.3_{-1.1}^{+2.2}\times10^{2}$&$1.6_{-1.4}^{+2.8}\times10^{1}$\\
NGC 1068  & XMM2 &$6_{-3}^{+4}\times10^{23}$ & $3.18_{-0.09}$ & $8.4_{-0.2}^{+0.3}\times10^4$
&$4_{-2}^{+3}\times10^{-2}$&$7_{-4}^{+5}$&$1.0_{-0.5}^{+0.7}$\\
NGC 6240  & XMM2 &$1.8\pm1.7\times10^{24}$ & $3.5_{-0.8}$  & $4.3_{-2.6}^{+1.0}\times10^4$
&$3.4_{-3.3}^{+3.8}$&$1.1\pm1.0\times10^{1}$&$8_{-8}^{+7}\times10^{-1}$\\
       & XMM4 &$1.8_{-1.7}^{+2.3}\times10^{24}$ & $3.1_{-0.3}^{+1.5}$    & $3.2_{-0.4}^{+0.7}\times10^4$
       &$4.6_{-4.6}^{+5.9}$&$1.1_{-1.1}^{+1.4}\times10^{1}$&$5.9_{-5.9}^{+7.6}\times10^{-1}$\\
\enddata
\tablecomments{Errors are quoted at the statistic 90\% level.
}
\end{deluxetable*}

Most UFOs are episodic; for example, in NGC 6240, UFO absorption lines are detected in XMM2 (exposure time is 4763~s) and XMM4 (10678~s) observations, whereas not in XMM1 (10119~s) or XMM3 (3050~s).
This episodicity may be due to change of the column density, the ionization state, and/or the wind geometry (e.g. \citealt{cap09}).
Here, we assume that the mass-loss rate is zero when the UFO line is not seen because its estimation is difficult without information of the absorption line.
This assumption provides us the lower limit of the mass-loss rate.
In NGC 6240 case, the ``time-averaged''  $v_{\rm UFO}$ is calculated to be $(4.3\times10^4\mathrm{\,km\,s^{-1}}\times4763\mathrm{\,s}+3.2\times10^4\mathrm{\,km\,s^{-1}}\times10678\mathrm{\,s})/(4763\mathrm{\,s}+10678\mathrm{\,s})=3.5\times10^4$~km~s$^{-1}$.
The average $\dot{M}_{\rm UFO}$ is calculated to be 
$(0\,M_\odot\mathrm{\,yr^{-1}}\times10119\mathrm{\,s}+3.4\,M_\odot\mathrm{\,yr^{-1}}\times4763\mathrm{\,s}+0\,M_\odot\mathrm{\,yr^{-1}}\times3050\mathrm{\,s}+4.6\,M_\odot\mathrm{\,yr^{-1}}\times10678\mathrm{\,s})/(10119\mathrm{\,s}+4763\mathrm{\,s}+3050\mathrm{\,s}+10678\mathrm{\,s})=2.3$~$M_\odot$~yr$^{-1}$.
The average $\dot{P}_{\rm UFO}$ and $\dot{K}_{\rm UFO}$ are also calculated in the same way.
These time-averaged parameters are listed in table \ref{tab:outflow}.

\begin{deluxetable*}{lccccc}
\tablewidth{0pt} 
\tablecaption{Properties of UFOs \label{tab:outflow}}
\tablehead{
\colhead{Object}&
\colhead{$v_{\rm UFO}$}&
\colhead{$\dot{M}_{\rm UFO}$}&
\colhead{$\dot{P}_{\rm UFO}$}&
\colhead{$\dot{K}_{\rm UFO}$}&
\colhead{$C$}\\
\colhead{}&
\colhead{(km s$^{-1}$)}&
\colhead{($M_\odot$~yr$^{-1}$)}&
\colhead{$(L_{\rm AGN}/c)$} & 
\colhead{($L_{\rm AGN}$)}&
\colhead{($=\dot{K}_{\rm mol}/\dot{K}_{\rm UFO}$)}
}
\startdata
 IC 5063  & $9.29_{-0.14}^{+0.13}\times10^4$ &  $1.3_{-0.6}^{+1.3}\times10^{-1}$&$1.2_{-0.5}^{+1.1}\times10^1$&$1.8_{-0.8}^{+1.7}$&$6_{-5}^{+6}\times10^{-3}$\\
I Zw 1 & $7.2\pm0.3\times10^4$&$2.4_{-1.8}\times10^{-2}$&$1.4_{-1.1}\times10^{-1}$&$1.6_{-1.3}\times10^{-2}$&$0.3^{+1.4}$\\
Mrk 231  & $6.3_{-0.3}^{+0.4}\times10^4$&$3.1_{-2.2}^{+4.6}\times10^{-2}$&$3.8_{-2.0}^{+5.6}\times10^{-2}$&$2.6_{-1.4}^{+4.7}\times10^{-3}$&$12_{-8}^{+13}$\\
Mrk 273  & $7.9\pm0.3\times10^4$&$3.7_{-3.2}^{+6.4}$&$1.0_{-0.9}^{1.8}\times10^2$&$1.3_{-1.2}^{+2.3}\times10^1$&$8_{-5}^{+2}\times10^{-3}$\\
NGC 1068 &$8.4_{-0.2}^{+0.3}\times10^4$&$1.8_{-0.9}^{+1.3}\times10^{-2}$&$3.4_{-1.7}^{+2.4}$&$4.8_{-2.4}^{+3.4}\times10^{-1}$&$4_{-2}^{+8}\times10^{-2}$\\
\,\,\,\,(ALMA) &&&&&$1.2_{-0.7}^{+1.3}\times10^{-2}$\\
NGC 6240 & $3.5_{-0.8}^{+1.0}\times10^4$&$2.3_{-2.3}^{+2.8}$&$5.9_{-5.9}^{+7.0}$&$3.5_{-3.5}^{+4.1}\times10^{-1}$&$6_{-3}\times10^{-2}$\\
\,\,\,\,(ALMA) &&&&&$2.0_{-1.1}\times10^{-2}$\\
\hline
IRAS F11119+3257&$7.6\pm0.3\times10^4$&1.5&$1.3_{-0.9}^{+1.7}$&$0.15$&$1.0-2.7\times10^{-2}$\\
\enddata
\tablecomments{Errors are quoted at the statistic 90\% level. Results of IRAS F11119+3257 are based on \citet{tom15}.}
\end{deluxetable*}

\begin{figure}
\includegraphics[width=65mm,angle=270]{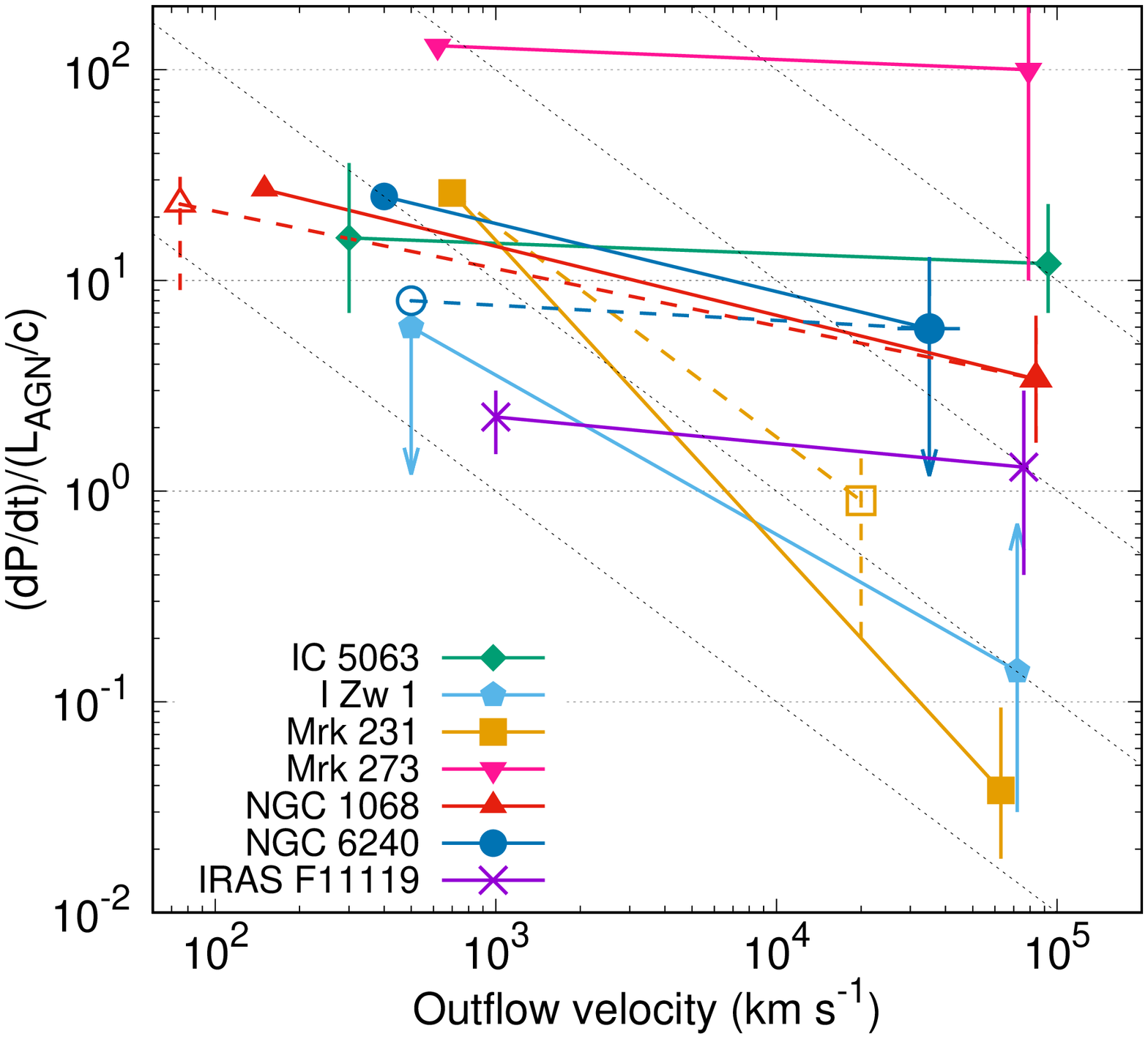}
\caption{Momentum versus outflow velocity.
The horizontal line shows the momentum-conserving flow, whereas the one ascending toward the left shows the energy-conserving flow.
The opened points of NGC 1068 (triangle) and NGC 6240 (circle) show the published ALMA results, and that of Mrk 231 (square) shows the published Chandra and NuSTAR result.
}
\label{fig:pv}
\end{figure}

Next, we compare the UFO parameters with those of the molecular outflows.
Figure \ref{fig:pv} shows momentum versus outflow velocity (also see figure 3 in \citealt {tom15} and figure 6 in \citealt{fer17}).
The horizontal line shows the momentum-conserving flow, whereas the one ascending toward the left shows the energy-conserving flow.
In this figure, Mrk 231 and I Zw 1 seem to be on the energy-conserving lines, whereas 
Mrk 273, IC5063 are on the momentum-conserving ones.
NGC 1068 and NGC 6240 are located between the two types of flows.

\begin{figure}
\includegraphics[width=65mm,angle=270]{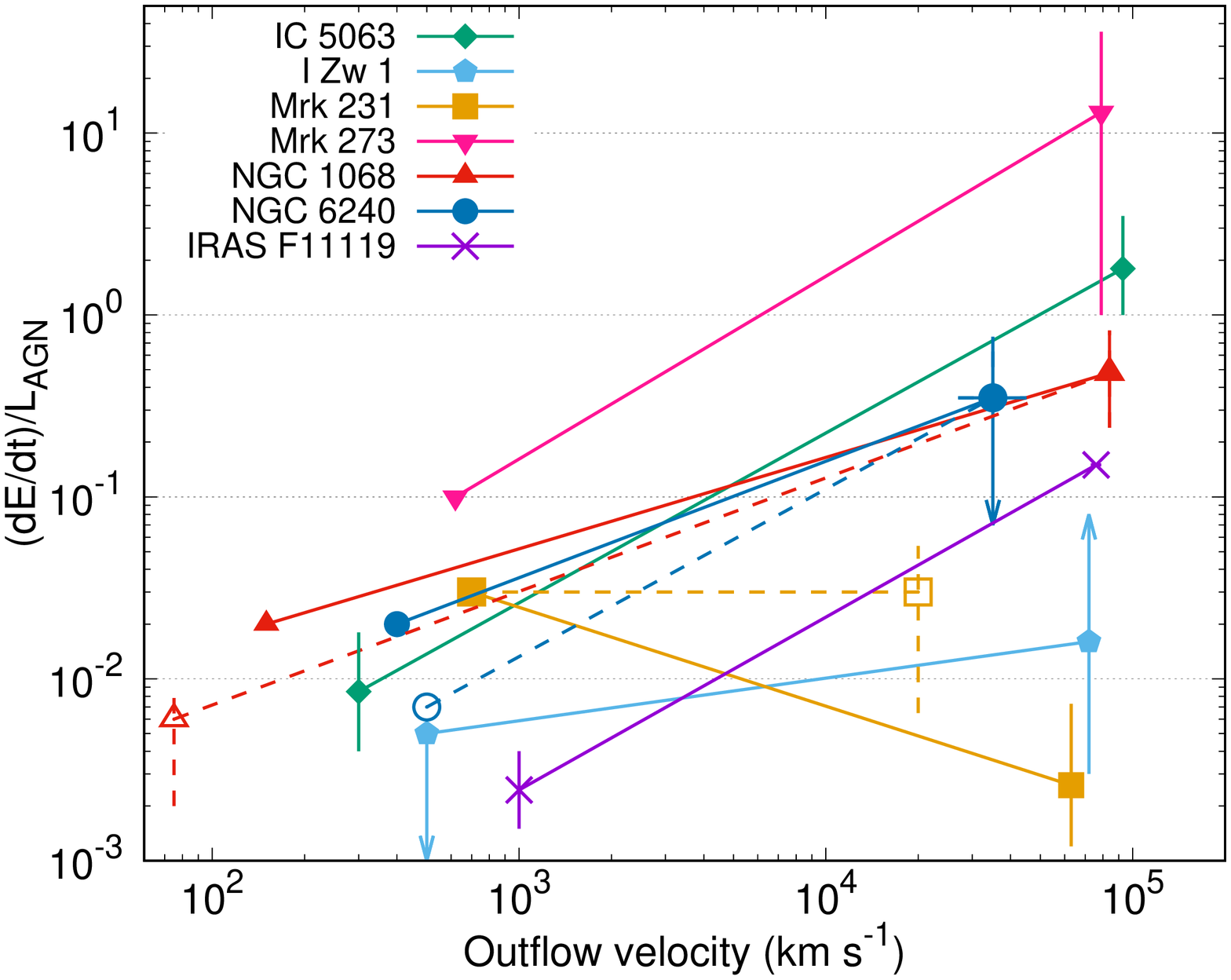}\\
\includegraphics[width=65mm,angle=270]{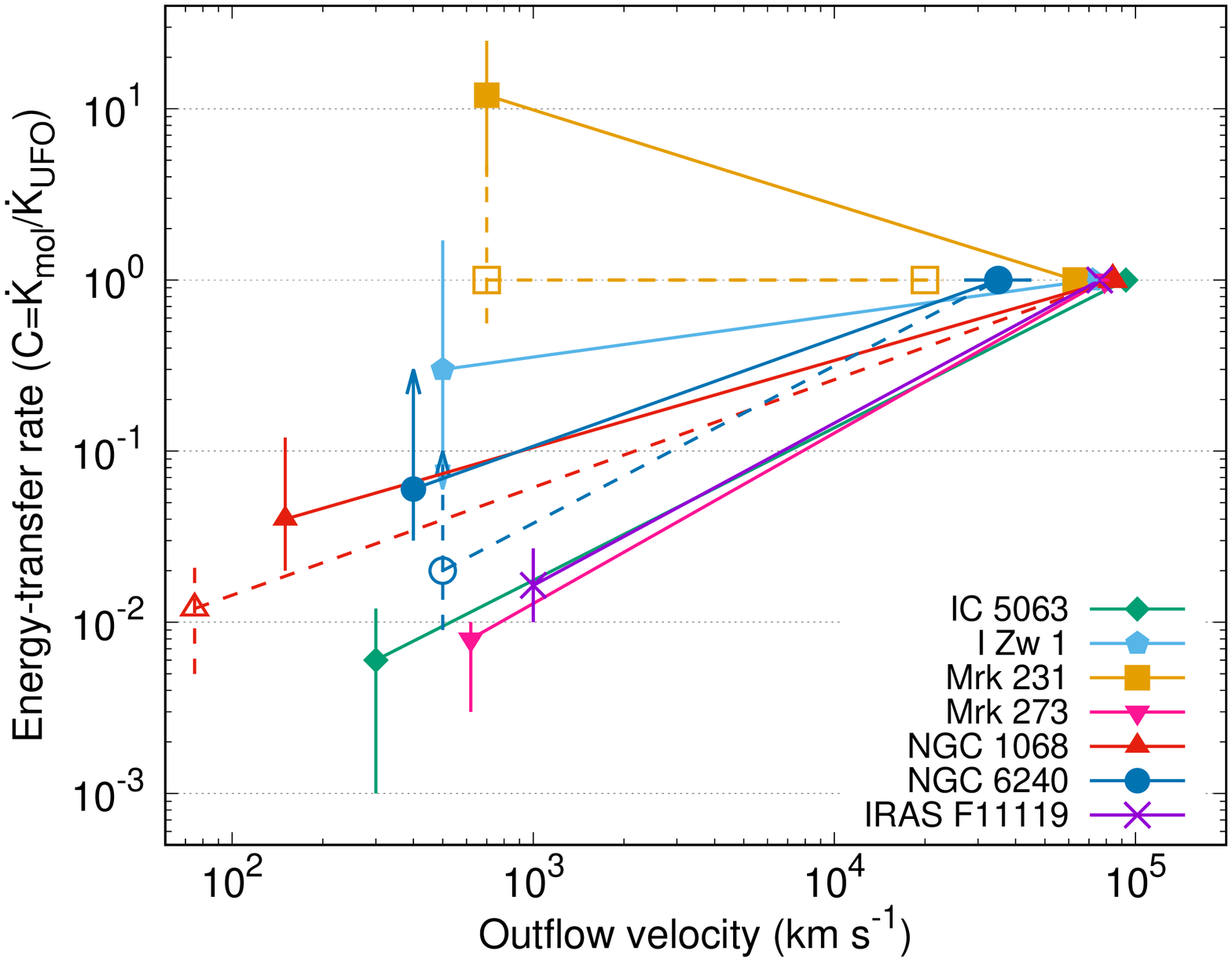}
\caption{Kinetic energy versus outflow velocity.
The vertical axis in the lower panel is normalized by the UFO kinetic energies, which corresponds to the energy-transfer rate $C$.
The point types are the same as in figure \ref{fig:pv}.}
\label{fig:ev}
\end{figure}

The upper panel of figure \ref{fig:ev} shows the kinetic energies versus the outflow velocities.
We can see that the kinetic energies in almost all the targets (except for Mrk 231) are lost 
between the two outflows,
which supports the idea that the molecular outflow is an accumulation of the ambient gas swept by the shock fronts UFOs have created and that radiative cooling occurs before the outflow gas reaches the shock front.
The energy-transfer rate of Mrk 231 exceptionally exceeds unity; this is probably because our data picked up weak UFOs.
\citet{fer15} analyzed the different data set of Mrk 231 (Chandra and NuSTAR observations), and derived the parameters of $v_{\rm UFO}=2.0_{-0.2}^{+0.3}\times10^4$~km~s$^{-1}$, $\dot{P}_{\rm UFO}/(L_{\rm AGN}/c)=0.2-1.6$, shown in the opened square point in figure \ref{fig:pv} and \ref{fig:ev}.
In this case, the data points are on the energy-conserving flow, i.e., $C$ is almost unity.
As a result, the energy-transfer rates are distributed between $C\sim0.007-1$ (see the bottom panel of figure \ref{fig:ev}).

\section{Discussion}\label{sec4}
\subsection{Parameter dependence}
\begin{figure}
\includegraphics[width=65mm,angle=270]{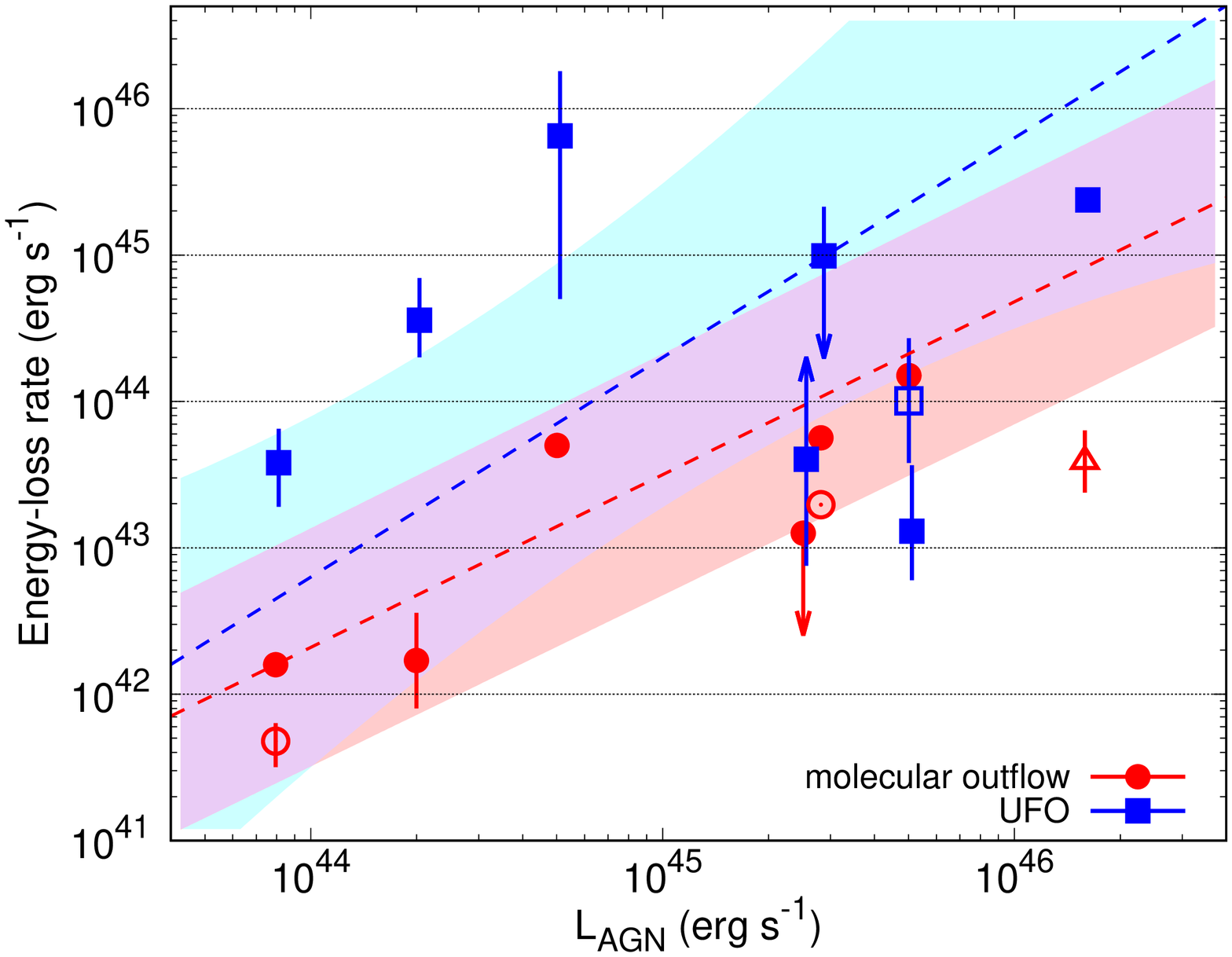}
\caption{Energy-loss rate ($\dot{K}$) versus AGN luminosity ($L_{\rm AGN}$). 
The blue/red points show the UFO/molecular outflow, respectively.
The red-filled circles show the IRAM data, whereas the red-open circles show the ALMA data.
The red triangle shows the Herschel data of IRAS F11119+3257.
The blue-filled squares show the results of this work, whereas the open one is the Chandra+NuSTAR data \citep{fer15}.
The dotted lines are the best-fit linear functions for larger samples in \citet{gof15} for UFO and \citet{cic14} for molecular outflows, whose error ranges are shown in the shaded areas.}
\label{fig:LAGN_K}
\end{figure}

\begin{figure}
\includegraphics[width=65mm,angle=270]{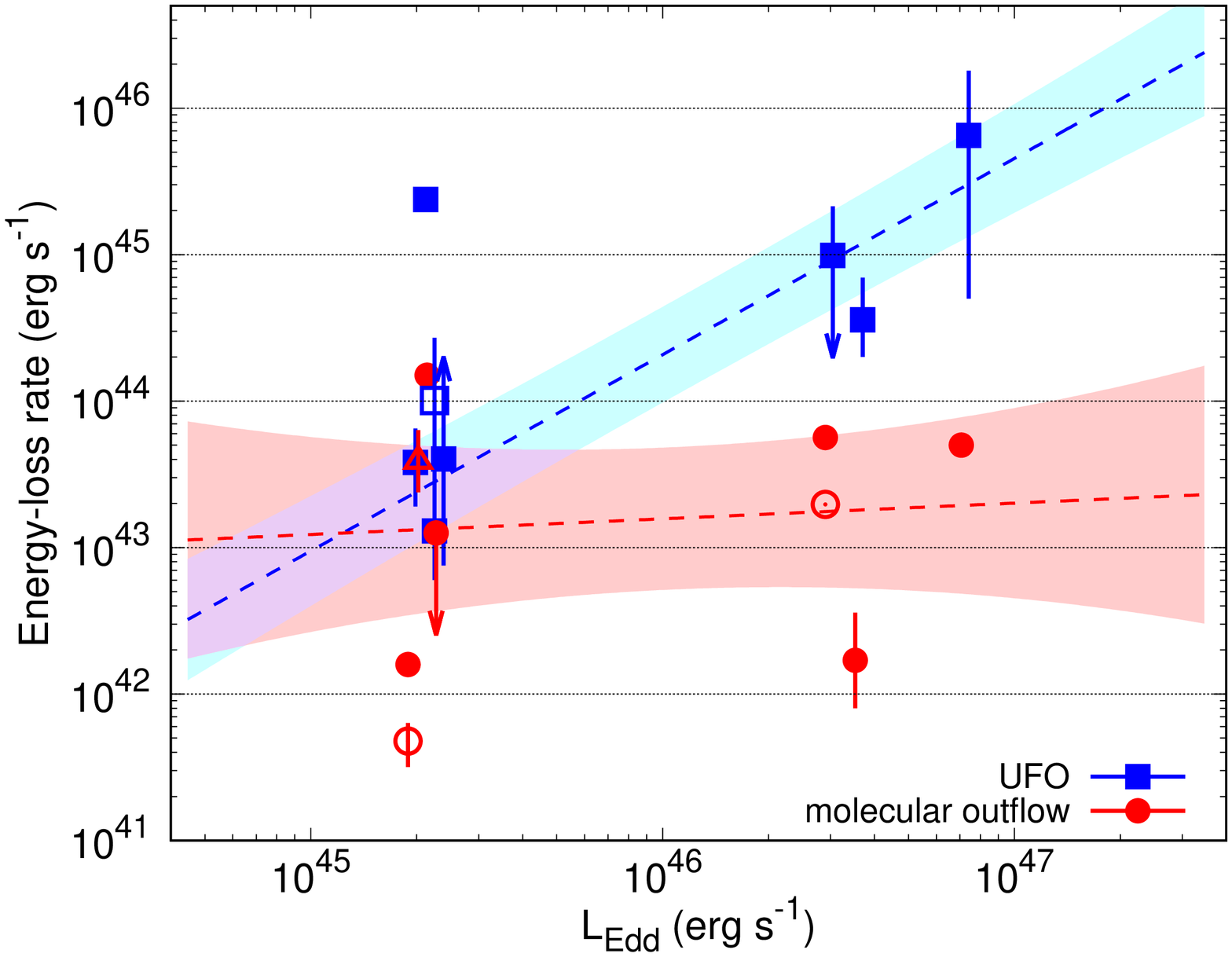}
\caption{Energy-loss rate ($\dot{K}$) versus Eddington luminosity ($L_{\rm Edd}$). See the caption of \ref{fig:LAGN_K} for details.}
\label{fig:LEdd_K}
\end{figure}

Figures \ref{fig:LAGN_K} and \ref{fig:LEdd_K} show $L_{\rm AGN}$- and $L_{\rm Edd}$ dependence of the energy-loss rates, respectively.
First, in figure \ref{fig:LAGN_K}, kinetic energies of each type of outflows are expected to have positive correlation to the AGN luminosities; $\log \dot{K}_{\rm UFO}=(-23.5^{+23.6}_{-44.7})+(1.5^{+1.0}_{-0.8})\log L_{\rm AGN}$ \citep{gof15} and $\log \dot{K}_{\rm mol}=(-9.6\pm6.1)+(1.18\pm0.14)\log L_{\rm AGN}$ \citep{cic14}.
Our samples are roughly on these correlations, and both the outflows seem to share the similar dependence for the AGN luminosities.
Next, in figure \ref{fig:LEdd_K}, we can see that the UFO kinetic energy has a positive correlation to the Eddington luminosities (i.e., BH masses).
This correlation can be explained by equation (\ref{eq1}), which means that larger energies are needed to escape stronger gravitational fields of heavier BH.
IRAS F11119+3257 has exceptionally strong kinetic energy (the isolated blue point in the top-left side of Fig.~\ref{fig:LEdd_K}), so we fit the data points except IRAS F11119+3257 with a linear function.
The best-fit function is $\log(\dot{K}_{\rm UFO})=44.32\pm0.30 + (1.34\pm0.23)(\log L_{\rm Edd}-46)$,
which is shown in the blue line in Fig.~\ref{fig:LEdd_K}).
On the other hand, the kinetic energies of molecular outflows are almost independent of the Eddington luminosities.
The best-fit linear function expect for IRAS F11119+3257 is
$\log(\dot{K}_{\rm mol})=43.19\pm0.38 + (0.11\pm0.57)(\log L_{\rm Edd}-46)$ (the red line).
This implies some feedback mechanism to suppress the kinetic energy of molecular outflows to a certain value, no matter how strong the UFO is.

\begin{figure}
\includegraphics[width=65mm,angle=270]{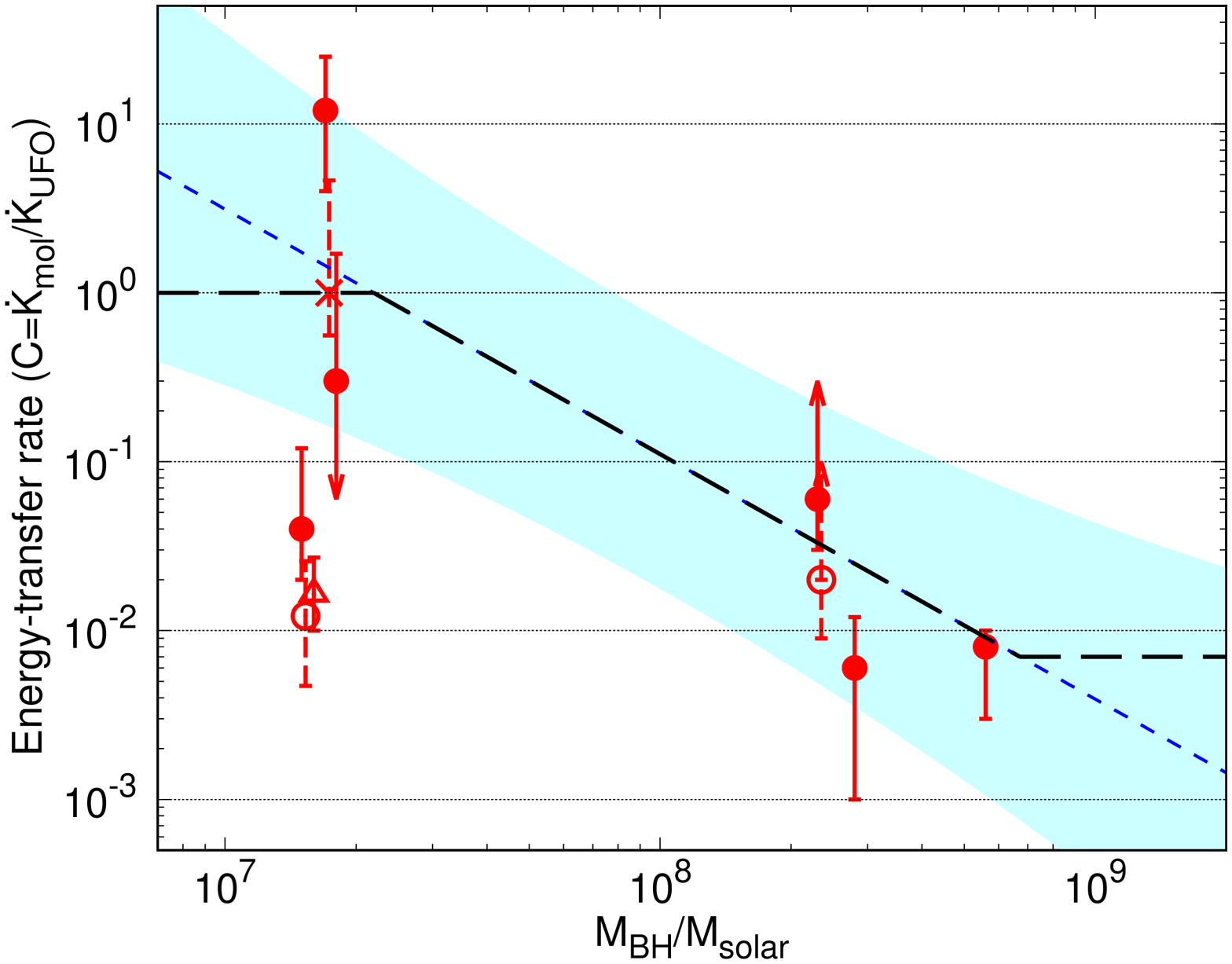}
\caption{Energy-transfer rate ($C$) versus BH mass ($M_{\rm BH}$). The filled circles show the IRAM data of the Seyfert galaxies, whereas the open circles show the ALMA data.
The cross bin is the Chandra+NuSTAR results of Mrk 231, and the open triangle shows the IRAS F11119+3257.
The blue line and the cyan-shaded region show the best-fit linear function and its error range, in which only the filled circle data points are used.
The black dashed line is the expected relation whose maximum is unity and minimum is $\sim0.007$.}
\label{fig:MBH_C}
\end{figure}

Figure \ref{fig:MBH_C} shows the energy-transfer rate versus BH masses.
We can see negative correlation in Figure \ref{fig:MBH_C}.
The energy-transfer rate reaches unity for the small BH masses, which means that the energy-conserving shock exists.
On the other hand, the momentum-conserving shock seems to exist when $M_{\rm BH}\gtrsim5\times10^8\,M_\odot$,
i.e., $C=v_{\rm molecular}/v_{\rm UFO}\sim500$~km~s$^{-1}/7\times10^4$~km~s$^{-1}\sim0.007$.
This is the minimum $C$ value.
The best-fit linear function is shown in the blue line in Fig.~\ref{fig:MBH_C}; 
$\log(C)=-0.96\pm0.64 + (-1.45\pm0.88)(\log M_{\rm BH} -8)$.
The black dashed line in Fig.~\ref{fig:MBH_C} is the expected correlation, in which $0.007\leq C\leq1$.
This negative correlation means that  the radiative cooling is more effective when the BH mass is larger.

\citet{kin03} said that whether the radiative cooling is effective or not depends on the balance of the cooling time scale of the outflowing gas and the flow time scale. 
The cooling efficiency of the outflowing gas depends on the balance of these two time scales \citep{kin03,kin11}.
Now we assume that the radius of the reverse shock between the unshocked UFO wind and the shocked UFO wind is small enough to be neglected and the hot bubble filled with the shocked UFO wind exists.
In this case, the hot bubble is thermalized and the Compton cooling may work.
\citet{kin03} shows that the Compton cooling time of the gas in the Eddington luminosity case is
\begin{equation}
t_{\rm cool}=\frac{2cR^2}{3\pi GM_{\rm BH}}\left(\frac{m_e}{m_p}\right)^2\left(\frac{v}{c}\right)^{-2}b,
\end{equation}
where $m_{e/p}$ is the electron/proton mass and $b(\lesssim1)$ is the filling factor for the collimation of the wind.
On the other hand, the flow time scale (for the momentum-driven case) is expressed as
\begin{equation}
t_{\rm flow}=R\left(\frac{2\pi G^2M_{\rm BH}}{f_{\rm gas}\sigma^2\kappa}\right)^{-1/2},
\end{equation}
where $f_{\rm gas}$ is the gas fraction to the dark matter, $\sigma$ is the velocity dispersion, and $\kappa$ is the opacity.
Consequently, the ratio of the two time scales is
\begin{equation}
\begin{split}
\frac{t_c}{t_f}
&=\frac{2}{3\pi}cR
\left(\frac{m_e}{m_p}\right)^2
\left(\frac{v}{c}\right)^{-2}b
\left(\frac{2\pi}{M_{\rm BH}f_{\rm gas}\sigma^2\kappa}\right)^{1/2}\\
&\simeq 1.8
\left(\frac{M_{\rm BH}}{10^8M_\odot}\right)^{-1/2}
\left(\frac{R}{1\,{\rm kpc}}\right) 
\left(\frac{v}{0.1c}\right)^{-2}.
\label{eq:balance}
\end{split}
\end{equation}
This equation shows that the cooling is more efficient (i.e., the energy-transfer rate is smaller) for larger BH masses, which is consistent with figure \ref{fig:MBH_C}.

\citet{ric18a,ric18b} performed the hydro-chemical simulations to demonstrate the molecular outflow swept by the inner outflow assuming UFO.
They isotropically injected wind particles within the inner boundary with the velocity of $0.1c$, assuming a spherically symmetric geometry.
They consider the radiative cooling in both the shocked UFO and the shocked ambient gas, and showed that 
the energy-transfer rate decreases in the higher BH masses
mainly due to stronger gravitational potential.
In the larger BH mass case, the velocity dispersion becomes larger and
the mass of the host galaxy enclosed within $R$, which is shown as $M_{\rm gal}(<R)=2\sigma^2 R/G$, becomes larger (see equation 2.2 in \citealt{ric18b}).
This tendency is also consistent with our results.

\begin{figure}
\includegraphics[width=65mm,angle=270]{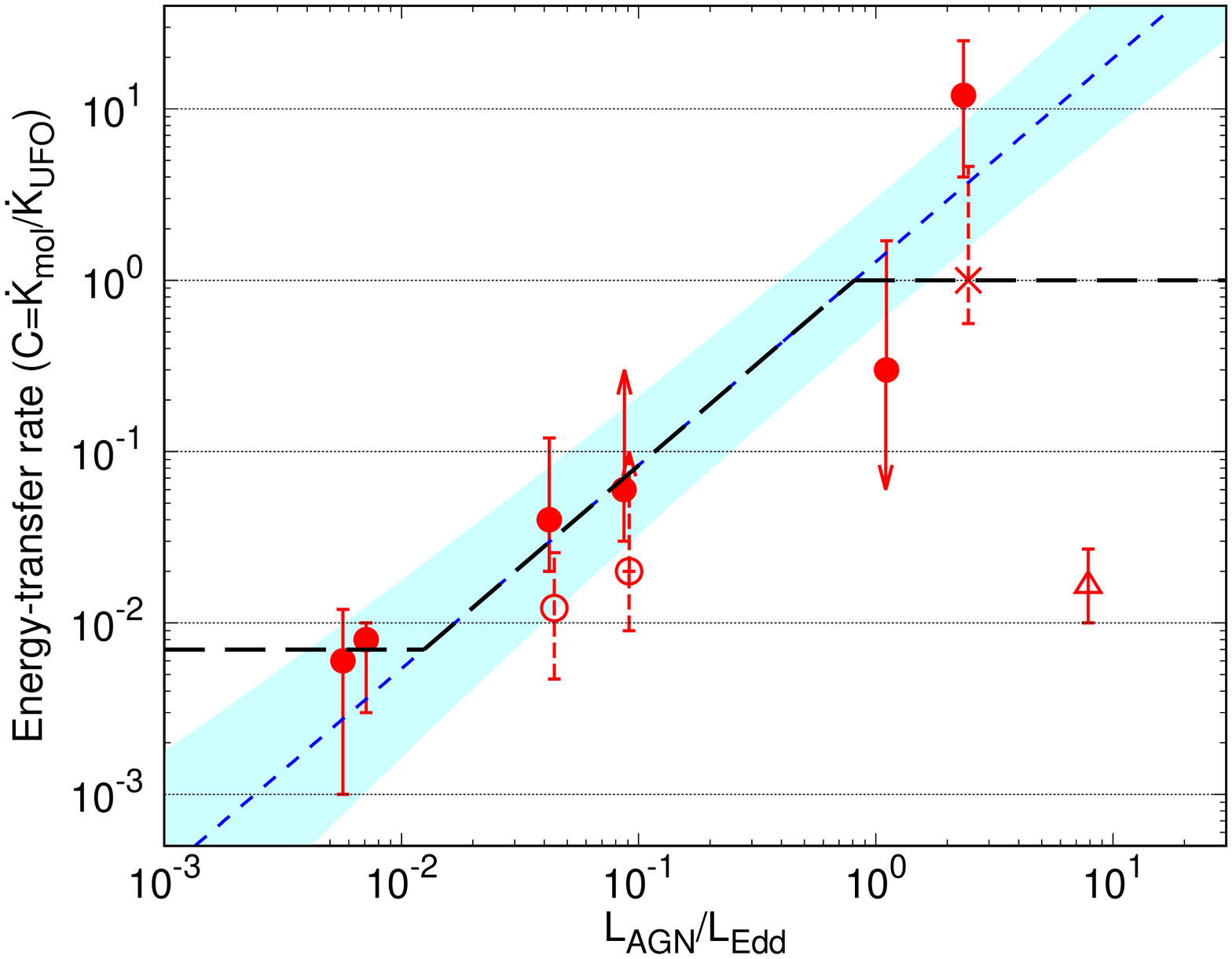}
\caption{Energy-transfer rate ($C$) versus Eddington ratios ($L_{\rm AGN}/L_{\rm Edd}$).
See the caption of figure \ref{fig:MBH_C} for details.}
\label{fig:edd}
\end{figure}

The other possibility is that the energy-transfer rate depends on the Eddington ratios.
From figures \ref{fig:LAGN_K} and \ref{fig:LEdd_K},
we can easily notice that the energy-transfer rate increases toward larger Eddington ratios (figure \ref{fig:edd}).
The energy-transfer rate reaches maximum at around the Eddington luminosity, and minimum when  $L_{\rm AGN}/L_{\rm Edd}\lesssim10^{-2}$.
The best-fit linear function is $\log(C)=0.11\pm0.28 + (1.19\pm0.33)\log(L_{\rm AGN}/L_{\rm Edd})$.
In this case, the quasar mode feedback is more efficient for Eddington/super-Eddington AGNs.
However, \citet{ric18b} shows that the energy-transfer rate is independent of AGN luminosity for the fixed $M_{\rm BH}=10^8M_\odot$,
which clearly contradicts our results.
Now the number of targets is very limited and the selection bias may exist;
our sample has a pseudo-correlation between $M_{\rm BH}$ and $L_{\rm AGN}/L_{\rm Edd}$.
More samples are needed to investigate the environmental dependence of AGN feedback more strictly.
If the energy-transfer rate is large for the larger Eddington ratios,
the BH mass may be fixed in the super-Eddington phase via strong accretion and strong feedback, because
most of the BH masses are considered to be acquired in the super-Eddington accretion phase \citep{kaw04}.

\subsection{Comments on uncertainty}
The energy-outflow rates of both the UFOs and molecular outflows have uncertainty.
The largest uncertainty in UFO parameters is the wind geometry, which determines $\Omega$ and $r$.
The X-ray reverberation lag techniques would make it possible to constrain $\Omega$ (see \citealt{miz18c}), but this method is not yet well established.
Ratios of the triplet lines in some ions (like Si and Fe) can constrain $n$ of the X-ray absorbers, but the current grating instrument can make only a rough constraint even with the good photon statistics (e.g., $n>10^7$~cm$^{-3}$ for NGC 5548; \citealt{mao17}); therefore $r$ is not yet well constrained.
In addition to it, the CCD calibration uncertainties have been reported;
absorption-line-like features are sometimes detected at $\sim9$~keV in the Crab data, which must have no intrinsic absorption lines in this energy band (see figure A2 in \citealt{kol14}).
This means that we may misdetect UFO lines in the ``featureless'' X-ray energy spectra.
Future missions with greater energy resolution and/or larger effective areas, such as X-Ray Imaging and Spectroscopy Mission (XRISM) and Advanced Telescope for High ENergy Astrophysics (Athena), will make it possible to detect the absorption features more confidently and let us know the detailed UFO parameters.
In molecular outflows, their sizes are most difficult to constrain.
Indeed, the IRAM observations in \citet{cic14} estimated size of the outflowing gases with simple modeling of their visibility, and cannot see their detailed geometry.
This estimation seems to overestimate the kinetic energies of molecular outflows by about three factors (see table \ref{tab:outflow}).
Therefore, spatially resolved observations with ALMA and IRAM (for nearby targets) are needed for more samples.
Consequently, for both UFOs and molecular outflows, an increasing number of samples with less uncertainty is required for detailed studies of energy transfer in outflows.

\section{Conclusion}\label{sec5}
To test whether the UFO kinetic energies are efficiently transferred into the galactic-scale molecular outflows and contribute to the AGN feedback,
we investigate the energy-transfer rate for larger samples.
The energy-transfer rate is defined as $C=\dot{K}_{\rm molecular}/\dot{K}_{\rm UFO}$, where $K_i$ is the kinetic energies of molecular outflows and UFOs.
We analyzed the X-ray (XMM-Newton and Suzaku) archive data of the targets that the molecular outflows are detected in IRAM/PdBI observations listed in \citet{cic14}, and
derived the energy-transfer rates for six Seyfert galaxies (plus type 1 quasar IRAS F11119+3257).
The energy-transfer rates are distributed between 0.007 (for the momentum-conserving shock) and 1 (for the energy-conserving shock).
We can see the correlation that the energy-transfer rate increases toward larger Eddington ratios (or lower BH masses), which can be explained by the balance of cooling time scale and flow time scale.
Consequently, we have found that UFO contribution to the AGN feedback is effective when the Eddington ratio is large and/or BH mass is small.

\acknowledgments
MM thanks Dr.~A.~J.~Richings, Prof.~C.~Done and Prof.~K.~Ohsuga for their comments and discussion.
The authors thank the data archive teams of XMM-Newton (ESA) and Suzaku (JAXA/ISAS).
MM is financially supported by JSPS Overseas Research Fellowships.
This work is supported by JSPS KAKENHI Grant Number 17K14247 (TI) and 17H06130 (KK).
The publication fee is supported by National Astronomical Observatory of Japan. 

\facilities{XMM, Suzaku, IRAM:Interferometer}
\software{HEASoft v6.23, SAS v15.0.0}

\bibliography{00}



\end{document}